\documentclass[useAMS,usenatbib]{mnras}
\usepackage{graphicx,amssymb,amsmath,times,verbatim,pdfpages,multirow,booktabs,array,dcolumn}
\usepackage[caption=false]{subfig}
\usepackage[flushleft]{threeparttable}

\title[IDV and PSD of 3C 273]{X-ray Intraday Variability and Power Spectral Density Profiles of the Blazar 3C 273 with {\it XMM-Newton} during 2000 -- 2021}
\author[Pavana Gowtami et al.]{G. S. Pavana Gowtami$^{1}$\thanks{Email: g.pavana.s.gowtami7@gmail.com}, Haritma Gaur$^{2}$\thanks{Email: harry.gaur31@gmail.com}, Alok C. Gupta$^{2}$\thanks{Email: acgupta30@gmail.com}\thanks{Corresponding Author}, Paul J. Wiita$^{3}$, Mai Liao$^{4,5}$, \\
\newauthor Martin Ward$^{6}$ \\
\\
$^{1}$Department of Physics, Indian Institute of Technology (IIT) Bombay, Powai, Mumbai -- 400076, India \\  
$^{2}$Aryabhatta Research Institute of Observational Sciences (ARIES), Manora Peak, Nainital -- 263001, India \\
$^{3}$Department of Physics, The College of New Jersey, 2000 Pennignton Rd., Ewing, NJ 08628-0718, USA \\
$^{4}$CAS Key Laboratory for Researches in Galaxies and Cosmology, Department of Astronomy, University of Science 
and Technology of China, Hefei, \\
~ Anhui 230026, China \\
$^{5}$School of Astronomy and Space Science, University of Science and Technology of China, Hefei, Anhui 230026, China \\
$^{6}$Centre for Extragalactic Astronomy, Department of Physics, University of Durham, South Road, Durham DH1 3LE, UK}

\begin{document}
\label{firstpage}
\pagerange{\pageref{firstpage}--\pageref{lastpage}}
\maketitle

\begin{abstract}
\noindent
\\
We present X-ray intraday variability and power spectral density (PSD) analyses of the longest 23 pointed {\it XMM-Newton} observations of the blazar 3C 273 that were taken during 2000 -- 2021. These good time intervals contained between 5 and 24.6 hours of data. Variability has been estimated in three energy bands: 0.2 -- 2 keV (soft), 2 -- 10 keV (hard), and 0.2 -- 10 keV (total). Nine of the 23 observations exhibited some variability, though no major variations exceeding 5 per cent were detected. Typical timescales for variability were $\sim 1$ ks. For those variable light curves we find that a  power-law model provides good fits to each PSD, with most of the slopes between $-1.7$ and $-2.8$. Although no variations of hardness ratio could be measured in any individual observation, an anti-correlation in flux and hardness ratio is found in long term data that indicates a harder when brighter trend. 
Our flux and spectral analyses indicate that both  particle acceleration and synchrotron cooling processes make an important contribution to the emission from this blazar.

\end{abstract}

\begin{keywords}
galaxies: quasars: general -- quasars: individual: 3C 273 -- radiation mechanisms: non-thermal 
\end{keywords}

\section{Introduction} 

\noindent
Super massive black holes (SMBHs) in the mass range $\sim 10^{6} - 10^{10} M_{\odot}$ are present at the center of active galactic nuclei (AGN)
which accrete matter from their surroundings and emit both thermal and non-thermal radiation. A large fraction of AGN, $\sim$ 85 - 90 \%, either emit very little or almost no radiation in radio bands and are called radio quiet (RQ) AGN, but the remaining $\sim$ 10 - 15 \% emit strongly in radio bands and are known as radio loud (RL) AGN \citep{1989AJ.....98.1195K}. A standard AGN classification is based on the importance of relativistic jet emission, in which RQAGN lack significant relativistic jets while  RLAGN have strong relativistic jets \citep{2017NatAs...1E.194P}. \\     
\\
Blazars are a subclass of RLAGN in which a powerful relativistic jet is closely aligned to the observer's line of site \citep{1995PASP..107..803U}. BL Lacertae objects and flat spectrum radio quasars (FSRQs) are collectively known as these blazars. In the composite optical/UV spectrum, BL Lacertae objects show either featureless or only very weak emission lines (equivalent width EW $\leq$ 5\AA) \citep{1991ApJ...374...72S,1996MNRAS.281..425M} while FSRQs have prominent emission lines \citep{1978PhyS...17..265B,1997A&A...327...61G}. In the present age of multi-wavelength (MW) transient astronomy, blazars are among one of the most well studied astronomical transients, because they emit radiation across the complete electromagnetic (EM) spectrum, their fluxes and polarizations are highly variable, and this emission is predominantly non-thermal. Since blazars emit radiation from the whole EM spectrum, their MW spectral energy distributions (SEDs) can be studied. Blazars' SEDs evince a double-hump structure in which the first hump (lower energy) peaks somewhere between infrared (IR) and X-ray energies and the emission is dominated by synchrotron radiation from the relativistic jet, while the second hump (higher energy) located in GeV to TeV $\gamma-$rays energies can be produced by various leptonic and/or hadronic-based emission processes \citep[e.g,][]{1998A&A...333..452K,2003APh....18..593M,2004NewAR..48..367K,2010ApJ...718..279G}.  \\
\\
Blazar flux variations in all EM bands range over all observable time scales ranging from a few minutes to several decades.  Flux variability detected in blazars on time scales of a few minutes to less than a day is often known as intra-day variability (IDV) \citep{Wagner1995}; variations in flux from days to a few months  are commonly known as short-term variability (STV); while the flux variations over longer timescales are often called long-term variability (LTV) \citep[e.g.,][]{2004A&A...422..505G}. \\
\\
3C 273\footnote{https://www.lsw.uni-heidelberg.de/projects/extragalactic/charts/1226+023.html} ($\alpha_{\rm 2000}$ = 12h 29m 06.698s; $\delta_{\rm 2000} = +02^{\circ} 03^{\prime} 08.58^{\prime\prime}$) at $z =$ 0.1575 is the first discovered quasar \citep{1963Natur.197.1040S}. It belongs to the FSRQ class of blazar and 
knots in its jet show apparent superluminal motion \citep{2001ApJS..134..181J,2005AJ....130.1418J}. 3C 273 has an additional ``big blue bump" (BBB) which dominates the optical through soft X-ray emission \citep{1998A&A...340...47P}. It is generally agreed that the BBB comes from more nearly isotropic emission from the close vicinity of the central SMBH, presumably from the accretion disc \citep{2002MNRAS.336..932K}. 3C 273 is among the most extensively studied blazars across the complete EM spectrum on diverse timescales. Observations have been carried out in single bands as well as in MW campaigns (simultaneous and/or non-simultaneous) \citep[e.g.,][and references therein]{2000A&A...354..513C,2000A&A...360...57R,2001ApJ...549L.161S,2003A&A...411L.343C,2005A&A...435..811P,2005ApJ...633L..85A,2006A&A...446...71S,2006ApJ...648..910U,2008A&A...486..411S,2009AJ....138.1428F,2010ApJ...714L..73A,2014ApJS..213...26F,Kalita2015,2016A&A...590A..61C,2017MNRAS.469.3824K,2019ApJ...880..155L,2020ApJ...903..134C,2020MNRAS.497.2066F}.    \\
\\
Focusing on X-ray observations, a $ \sim$18 ks integration of 3C 273 in January 1992 with {\it ROSAT HRI}, clearly detected a X-ray jet  all along the optically visible jet as well as a faint X-ray halo with a characteristic scale of 29 kpc \citep{2000A&A...360...99R}. \citet{2001ApJ...549L.161S} reported {\it Chandra} observations of the X-ray jet of 3C 273 during the calibration phase in 2000 January. They detected the brightest optical knots in the 0.2 -- 8 keV energy band and found that the X-ray morphology nicely tracks that of the optical. By using five observations of 3C 273 with {\it Chandra}, at least four distinct features were resolved in the jet, and it was found that the X-ray emission  mostly arises from the ``inner jet" between 5$^{\prime \prime}$ and 10$^{\prime \prime}$ from the core \citep{2001ApJ...549L.167M}.  A deep {\it Chandra} observation of the high-powered FSRQ jet, along with radio and optical measurements, showed that the X-ray spectra were much softer than the radio spectra in all regions of the bright part of the jet, except for the first bright ``knot A". These observations ruled out a model in which the X-ray emission from the entire jet arises from beamed inverse Compton (IC) scattering of cosmic microwave background photons in a single-zone jet flow \citep{2006ApJ...648..900J}.   It has been suggested  that the X-ray and $\gamma-$ray fluxes from ``knot A" may have a common origin in synchrotron emission from the accelerated protons \citep{2014MNRAS.444L..16K}. Combined temporal and spectral analyses of 3C 273 suggest that a two-component model is needed to explain the complete high energy spectrum, where X-ray emission is likely to be dominated by a Seyfert-like component while the $\gamma-$ray emission is dominated by a blazar-like component produced by the relativistic jet \citep{2015A&A...576A.122E}. \\
\\
The FeK$\alpha$ line is an important physical diagnostic in X-ray spectra of AGNs and quasars but has been extremely difficult to measure in the blazar 3C 273. Nonetheless, it has been detected occasionally in 3C 273. \citet{2000ApJ...544L..95Y} detected strong Fe K$\alpha$ line from a 1996 July observing campaign of 3C 273 with {\it RXTE} and {\it ASCA}, and another detection could be made from all the data on the blazar taken by {\it ASCA} from 1993 -- 2000. In a {\it XMM-Newton} observation taken on 7 July 2003 (Obs ID 159960101, see Table 1) the FeK$\alpha$ line also was  detected \citep{2017MNRAS.469.3824K}. The FeK$\alpha$ line is generally seen in Seyfert 1 galaxies from which it can be concluded that the X-ray  emission from 3C 273 is a mixture of  a thermal AGN component associated with the inner accretion flow emission from the disc's corona as well as non-thermal processes produced in the jet component \citep{1995MNRAS.273..837M}. \\
\\
Some of the most puzzling flux variations are those happening on IDV timescales. Study of IDV is an important method for learning about structures on small spatial scales, and also provides us with better understanding of the different radiation mechanisms that are important in the emitting regions in the vicinity of the central engine of blazars \citep{Wagner1995}. To better understand blazars' flux variabilities on IDV timescales, over the past decade and more, we have been  using public archive data of blazars taken from various X-ray satellites (e.g. {\it XMM-Newton, NuStar, Chandra}, and {\it Suzaku}). We reported these results in a series of papers  \citep{2009A&A...506L..17L,2010ApJ...718..279G,2014MNRAS.444.3647B,2016NewA...44...21B,Kalita2015,2016MNRAS.462.1508G,Pandey2017,2018ApJ...859...49P,2018MNRAS.480.4873A,2019ApJ...884..125Z,2021ApJ...909..103Z,2021MNRAS.506...1198D}. In a sample of 24 {\it XMM-Newton} LCs of 4 high energy peaked blazars (PKS 0548$-$322, ON 231, 1ES 1426$+$428, and PKS 2155$-$304), a $\sim$4.6 hour quasi periodic oscillation (QPO) was detected in a 0.3 -- 10 keV LC of PKS 2155$-$304 \citep{2009A&A...506L..17L}, and possible weak QPOs may have been present in LCs of ON 231 and PKS 2155$-$304. IDV timescales ranging from 15.7 to 46.8 ks were present in eight of their LCs \citep{2010ApJ...718..279G}. 
By using 20 {\it XMM-Newton} pointed observations of PKS 2155$-$ 304 with simultaneous X-ray and UV/optical data, spectral energy distributions (SEDs) were constructed and fitted with a combined power-law and log-parabolic model  \citep{2014MNRAS.444.3647B}. 
Three continuous pointings of $\sim$ 92 ks of PKS 2155$-$304 on 24 May 2002 with {\it XMM-Newton} displayed a mini-flare, a nearly constant flux period and a strong flux increase \citep{2016NewA...44...21B}. 
Two dozen pointed X-ray observations of the low energy peaked blazar 3C 273 taken during 2000 -- 2012 with {\it XMM-Newton} displayed occasionally very low amplitude flux variation \citep{Kalita2015} and here we expand upon that work. \\
\\
In an extensive X-ray IDV study of the 12 low energy peaked blazars involving 50 observations from {\it XMM-Newton} since its launch to 2012, it was found that this class is not very variable in these X-ray bands, with a duty cycle $\sim$ 4 per cent \citep{2016MNRAS.462.1508G}. An examination of 46 LCs taken with the {\it Nuclear Spectroscopic Telescope Array (NuSTAR)} of 11 TeV emitting blazars, found that 6 of these blazars exhibited IDV in the NuSTAR energy range of 3 -- 79 keV \citep{Pandey2017,2018ApJ...859...49P}. An extensive study of 72 {\it Chandra} LCs of the TeV blazar Mrk 421 between 2000 and 2015, showed  that this source often displayed IDV in the energy range 0.3 -- 10.0 keV with duty cycle of $\sim$ 84 per cent \citep{2018MNRAS.480.4873A}. In 16 pointed observations of the TeV blazars Mrk 421 and PKS 2155$-$304 taken during the whole operational period of {\it Suzaku}, large amplitude IDV was seen every time \citep{2019ApJ...884..125Z,2021ApJ...909..103Z}.  In a recent study of twenty pointed {\it XMM-Newton} observations of the TeV blazar PG 1553+113 taken during 2010 -- 2018, IDV was seen in the X-ray energy range (0.3 -- 10 keV) in 16 out of 19 LCs, or a duty cycle of $\sim$ 84\% \citep{2021MNRAS.506...1198D}. \\
\\
In blazars, IDV across the complete EM spectrum is usually dominated by radiation changes intrinsic  to the source, except for  low-frequency radio observations that may include fluctuations of extrinsic origin arising from interstellar scintillation \citep[e.g.][]{Wagner1995}. The bulk of the intrinsic flux variability in blazars at all wavelengths, including X-rays, can be explained by the relativistic shock-in-jet based radio loud AGN emission models, particularly those that include relativistic turbulence or magnetic reconnection in the jets that can produce the fast IDV \citep[e.g.,][and references therein]{1985ApJ...298..114M,1992A&A...259..109G,2014ApJ...780...87M,2015JApA...36..255C,2016ApJ...820...12P,2021MNRAS.502.1145Z,2021A&A...649A.150B}. In other classes of AGNs and for blazars in low-flux states, variations in optical and UV emission can be explained with  accretion disc models \citep[e.g.,][]{1993ApJ...406..420M,1993ApJ...411..602C} where the X-rays arise from the disc corona \citep{2008A&A...486..411S,2017MNRAS.464.3194B,2021ApJ...910..103L}.  In the case of 3C~273, it has long been argued that there are substantial contributions from both the disc and the jet in both optical \citep{1989ApJ...347...96I,1998A&A...340...47P} and X-ray bands \citep{2004Sci...306..998G}.   Recently, \citet{2020ApJ...897...18L} have shown that the jet contribution to the optical emission of 3C 273 ranges between 10 and 40 per cent, with a mean of $\sim$ 28 per cent. \\
\\
For the present work, we have taken public archive data of the blazar 3C 273 taken by the EPIC-pn instrument on board the {\it XMM-Newton} satellite. There are a total of 23 pointed observations which have qualified under our selection criteria described in Section 2. These observations were carried out for a period spanning just over 21 years (2000 -- 2021). The data we discuss here are very useful to study flux variability, power density spectra and spectral variations on IDV timescales. This study should help us to understand the X-ray properties of this blazar on the smallest physical scales. \\
\\
The paper is arranged as follows. In Section 2, we discuss the {\it XMM-Newton} public archive data of the blazar 3C 273 and how we reduced it. Section 3 provides brief descriptions of the various analysis techniques we have used. In Section 4 the results are presented. We present a discussion and conclusions in sections 5 and 6, respectively.

\begin{table*}
{\bf Table 1.} Observation log of {\it XMM-Newton} X-ray data for 3C 273.
\centering
\noindent
\scalebox{1.2}{
\begin{tabular}{c c c c c c c c c c c} \hline
\hline
        &                &            &           &          & \multicolumn{3}{c} {$\mu$(counts/s)$^{b}$} &  \\\cline{6-8}
        &   Date of Obs. &            & GTI$^{a}$ &          & Soft         & Hard         & Total        & Mean \\
 ObsID  &   yyyy.mm.dd   & Revolution & (ks)      & Pileup   & (0.2-2 keV)  & (2-10 keV)   & (0.2-10 keV) & HR  \\ \hline
0126700301 & 2000.06.13 & ~~94 &  64.9  & no  &  ~42.16$\pm$0.83  & ~9.88$\pm$0.40 &  ~51.93$\pm$0.92 & $-$0.62$\pm$0.01 \\
0126700601 & 2000.06.15 & ~~95 &  29.6  & no  &  ~40.53$\pm$0.81  & ~9.68$\pm$0.40 &  ~50.10$\pm$0.90 & $-$0.61$\pm$0.01 \\
0126700701 & 2000.06.15 & ~~95 &  29.9  & no  &  ~39.32$\pm$0.80  & ~9.43$\pm$0.39 &  ~48.65$\pm$0.89 & $-$0.61$\pm$0.01 \\
0126700801 & 2000.06.17 & ~~96 &  56.5  & no  &  ~39.53$\pm$0.80  & ~9.43$\pm$0.39 &  ~48.85$\pm$0.89 & $-$0.61$\pm$0.01 \\
0136550101 & 2001.06.13 & ~277 &  88.6  & no  &  ~55.68$\pm$0.95 & 11.75$\pm$0.44 &  ~67.30$\pm$1.05 & $-$0.65$\pm$0.01 \\
0159960101 & 2003.07.07 & ~655 &  58.1  & yes &  ~86.87$\pm$3.99  & 15.67$\pm$1.70 &  102.54$\pm$4.34 & $-$0.69$\pm$0.03 \\
0136550801 & 2004.06.30 & ~835 &  18.0  & no  &  ~39.87$\pm$0.81  & ~8.34$\pm$0.37 &  ~48.12$\pm$0.89 & $-$0.65$\pm$0.01 \\
0136551001 & 2005.07.10 & 1023 &  27.6  & no  &  ~42.97$\pm$0.82  & 10.05$\pm$0.40 &  ~52.92$\pm$0.91 & $-$0.62$\pm$0.01 \\
0414190101 & 2007.01.12 & 1299 &  76.6  & no  &  ~45.82$\pm$0.87  & 14.76$\pm$0.50 &  ~60.42$\pm$1.00 & $-$0.51$\pm$0.01 \\
0414190301 & 2007.06.25 & 1381 &  32.0  & no  &  ~38.34$\pm$0.78  & 10.41$\pm$0.41 &  ~48.64$\pm$0.88 & $-$0.57$\pm$0.01 \\
0414190401 & 2007.12.08 & 1465 &  35.4  & no  &  ~81.05$\pm$1.15  & 19.37$\pm$0.57 &  100.20$\pm$1.28 & $-$0.61$\pm$0.01 \\
0414190501 & 2008.12.09 & 1649 &  40.5  & yes &  115.32$\pm$2.14  & 30.12$\pm$1.12 &  144.31$\pm$2.40 & $-$0.59$\pm$0.01 \\
0414190601 & 2009.12.20 & 1837 &  31.4  & no  &  ~62.76$\pm$1.01  & 15.43$\pm$0.51 &  ~78.01$\pm$1.13 & $-$0.61$\pm$0.01 \\
0414190701 & 2010.12.10 & 2015 &  35.9  & no  &  ~46.32$\pm$0.86  & 11.07$\pm$0.42 &  ~57.27$\pm$0.95 & $-$0.61$\pm$0.01 \\
0414190801 & 2011.12.12 & 2199 &  42.8  & no  &  ~43.31$\pm$0.84  & ~9.82$\pm$0.40 &  ~53.02$\pm$0.93 & $-$0.63$\pm$0.01 \\
0414191001 & 2012.07.16 & 2308 &  25.5  & no  &  ~37.50$\pm$0.78  & ~8.58$\pm$0.37 &  ~45.99$\pm$0.86 & $-$0.63$\pm$0.01 \\
0414191101 & 2015.07.13 & 2856 &  70.8  & no  &  ~32.33$\pm$0.72  & ~7.23$\pm$0.34 &  ~39.48$\pm$0.80 & $-$0.63$\pm$0.02 \\
0414191201 & 2016.06.26 & 3031 &  65.6  & no  &  ~53.86$\pm$0.93  & 14.98$\pm$0.49 &  ~68.69$\pm$1.05 & $-$0.56$\pm$0.01 \\
0414191301 & 2017.06.26 & 3214 &  65.4  & no  &  ~29.38$\pm$0.69  & ~7.85$\pm$0.36 &  ~37.14$\pm$0.78 & $-$0.58$\pm$0.01 \\
0414191401 & 2018.07.04 & 3401 &  63.1  & no  &  ~23.33$\pm$0.62  & ~6.65$\pm$0.33 &  ~29.91$\pm$0.70 & $-$0.56$\pm$0.02 \\
0810820101 & 2019.07.02 & 3583 &  67.5  & no  &  ~21.78$\pm$0.60  & ~5.83$\pm$0.31 &  ~27.54$\pm$0.67 & $-$0.58$\pm$0.02 \\
0810821501 & 2020.07.06 & 3768 &  67.9  & no  &  ~24.06$\pm$0.63  & ~7.05$\pm$0.34 &  ~31.03$\pm$0.71 & $-$0.55$\pm$0.02 \\
0810821601 & 2021.06.09 & 3938 &  57.7  & no  &  ~17.86$\pm$0.54  & ~4.89$\pm$0.28 &  ~22.69$\pm$0.61 & $-$0.57$\pm$0.02 \\\hline
\end{tabular}} \\
{\bf Notes:} $^{a}$ GTI = good time interval, $^{b}$ $\mu$ = mean count rate
\end{table*}

\vspace*{-0.5cm}
\section{{\it XMM-Newton} Archival Data Selection and Reduction}

\subsection{Data selection parameters}
In this work we consider the  FSRQ 3C 273, which was continuously monitored for  extended periods of time by the {\it XMM-Newton} satellite on many occasions since its launch until June 2021. We took only the X-ray data from the European Photon Imaging Camera (EPIC)-pn detector from the online {\it XMM-Newton} public archive\footnote{HEASARC (High-Energy Astrophysics Science Archive Research Centre)}. We did not incorporate EPIC-MOS (Metal Oxide Silicon) instrument data for the present study. The path of EPIC-MOS instruments is partially obscured by the RGS (Reflection Grating Spectrometer) instruments for focussing the incoming photons of the interested source so that the effective area of both EPIC-MOS 1 and MOS 2 is less than that of the EPIC-pn. In addition, the effective area of the EPIC-MOS decreases rapidly with respect to the EPIC-pn instrument above an energy of $\sim$ 4 keV. Another important reason for preferring the  EPIC-pn data is that has timing resolution almost 50 times higher compared to EPIC-MOS\footnote{https://xmm-tools.cosmos.esa.int/external/xmm\_user\_support/documentation/uhb/effareaonaxis.html}. We therefore preferred EPIC-pn data sets to study the variability properties of the object.  We note that for similar earlier studies, we have only used EPIC-pn data \citep[e.g.,][and references therein]{2010ApJ...718..279G,2014MNRAS.444.3647B,2016MNRAS.462.1508G,2021MNRAS.506...1198D}. \\
\\
The data, which consist of 43 pointed observations, were taken from 13 June 2000 to 9 June 2021, spanning almost exactly 21 years. We immediately eliminated 11 observations of which 1 does not have PN data while another 10 have few data points. Nine more observations are excluded as they have a good time interval (GTI) of less than 18 ks, or 5 hours. Our analysis is performed on the 23 remaining Obs IDs. A summary of the 3C 273 public archive data we used is given in Table 1. 

\subsection{Data Reduction}
The onboard system of {\it XMM-Newton} satellite consists of two CCDs, the MOS and PN cameras. As noted above, only the EPIC-pn data are considered here as they have higher count rates and more high energy sensitivity  than MOS data; they also have less pile-up distortion \citep[e.g.][]{Gonzales2012}. The EPIC-pn camera takes the image of the target in the energy range of 0.15 -- 15 keV in the X-ray band. However, the data above 10 keV is dominated by strong proton flaring. The on-axis effective area of the EPIC-pn camera reflects X-ray photons most efficiently at 0.2 to 10 keV. So to obtain high quality datasets, we consider the 0.2 -- 10 keV X-ray energy range.  \\
\\
We processed this data using the standard procedure of the {\it XMM-Newton} Science Analysis System (SAS) version 19.0.0 with the most recent available Current Calibration File\footnote{http://www.cosmos.esa.int/web/xmm-newton/sas-threads (I. de la Calle)}. 
The event files of the PN detector were generated through \emph{epproc}. Before generating a cleaned event list, we first examined the 
light curve (LC) in the energy range 10 -- 12 keV to find the  soft proton flares. To remove them, a good time interval (GTI) file using the \emph{tabgtigen} tool is generated which contains information of the time intervals free from these flares. In the next step, we used the event list file and GTI files as input to obtain cleaned event files using the \emph{evselect} tool. Then, filtering of the data is done using the condition (PATTERN $\leq 4$) and (FLAG $= 0 $) in 3 different energy bands: total (0.2 -- 10 keV), soft (0.2 -- 2 keV) and hard (2 -- 10 keV). The source counts are normally taken from a circular aperture of radius of 40 arcseconds and the background counts were found by taking a circular region that ranged between 40 and 50 arcsec radius for different images/CCD chips, always as far away as possible from the source on the same CCD chip.  
Any pile-up is detected using the \emph{epatplot} routine. The pile-up is removed by selecting a central annulus region instead of circular region for the source event file. All observations are binned at 100 seconds. By subtracting the background counts from the source counts, the final corrected events are obtained using the \emph{epiclccorr} task. High background periods at the beginning and end of some of the observations were removed to obtain the final corrected lists. \\ 
\\
The 23 publicly archived pointed {\it XMM-Newton} observations of the FSRQ blazar 3C 273 we analyzed have individual GTIs between 18.0 and 88.6 ks and were taken over an $\sim$21 year time span: the earliest pointed observation  was taken on 2000 June 13 and the last one on 2021 June 9. These  observations gave us an excellent opportunity to study the flux and spectral variability of the FSRQ blazar 3C 273 on IDV timescales over an extended period. The LCs in all the three bands are plotted; examples are shown in Figs.\ 1 and  2 and the all LCs are in the online Supplemental Material. Few of these LCs show obvious significant variability over these timescales, which range from 5 to 24.6 hr, so they must be analysed carefully to explore any variations.   

\begin{table*}
{\bf Table 2.} X-Ray variability parameters in soft, hard and total bands of 3C 273.

\centering
\noindent
\scalebox{0.8}{
\begin{tabular}{l c c c c c c c c c c c c c} \hline \hline
Observation ID  & \multicolumn{9}{c} {$F_{var}$ (percent)}  & ${|\tau|}^{c} (ks) $ &${|\tau|}^{d}_{corr} (ks)$ &$ |\tau|^{c} (ks)$ &$ |\tau|^{c} (ks)$   \\\cline{2-10}
       &             Soft      & Sig$^{a}$ &  Var$^{b}$&    Hard     & Sig$^{a}$ & Var$^{b}$  &   Total & Sig$^{a}$ & Var$^{b}$ & Total &  Total &  Soft &  Hard        \\
       &         (0.2-2 keV)   &           &           &  (2-10 keV) &             &          &   (0.2-10 keV) & & &  (0.2-10 keV) &  (0.2-10 keV) & (0.2-2 keV) & (2-10 keV)        \\ \hline 
0126700301 & 0.85$\pm$0.15 & 5.67 &  ~V & 0.17$\pm$2.74 & 0.06 &  NV & 0.64$\pm$0.15 & ~4.21 & NV &  $-$    &  $-$ & 1.06$\pm$0.31  &  $-$ \\
0126700601 & 0.19$\pm$0.88 & 0.22 &  NV & 0.25$\pm$2.81 & 0.09 &  NV & 0.28$\pm$0.48 & ~0.59 & NV       & $-$          & $-$           &       $-$          &       $-$ \\
0126700701 & 0.65$\pm$0.28 & 2.32 &  NV & 0.80$\pm$0.92 & 0.87 &  NV & 0.73$\pm$0.21 & ~3.43 & NV & $-$          & $-$           &       $-$          &       $-$ \\
0126700801 & 0.94$\pm$0.16 & 5.88 &  ~V & 1.46$\pm$0.40 & 3.65 &  NV & 0.95$\pm$0.13 & ~7.31 & ~V       & 1.27$\pm$0.41 & 1.09$\pm$0.35 & 1.16$\pm$0.38 & $-$ \\
0136550101 & 0.87$\pm$0.10 & 8.70 &  ~V & 0.43$\pm$0.79 & 0.54 &  NV & 0.71$\pm$0.10 & ~7.31 & ~V & 1.48$\pm$0.48 & 1.28$\pm$0.42 & 1.40$\pm$0.48 & $-$     \\
0159960101 & 1.30$\pm$0.52 & 2.50 &  NV & 1.49$\pm$2.39 & 0.62 &  NV & 1.12$\pm$0.50 & ~2.23 & NV & $-$  & $-$  &  $-$                     &       $-$                     \\
0136550801 & 0.85$\pm$0.29 & 2.93 &  NV & 1.42$\pm$0.81 & 1.75 &  NV & 0.84$\pm$0.25 & ~3.31 & NV &  $-$          & $-$           &       $-$          &       $-$ \\
0136551001 & 0.38$\pm$0.43 & 0.88 &  NV & 0.24$\pm$2.78 & 0.09 &  NV & 0.25$\pm$0.51 & ~0.49 & NV &$-$  & $-$  &    $-$                     &       $-$                     \\
0414190101 & 0.92$\pm$0.12 & 7.67 &  ~V & 1.82$\pm$0.20 & 9.10 &  ~V & 1.19$\pm$0.08 & 14.10 & ~V & 1.40$\pm$0.45 & 1.21$\pm$0.39 & 1.01$\pm$0.27 & 0.70$\pm$0.23   \\
0414190301 & 0.75$\pm$0.25 & 3.00 &  NV & 0.46$\pm$1.33 & 0.35 &  NV & 0.61$\pm$0.23 & ~2.65 & NV & $-$ & $-$  & $-$  & $-$                     \\
0414190401 & 0.47$\pm$0.18 & 2.62 &  NV & 0.51$\pm$0.65 & 0.78 &  NV & 0.51$\pm$0.14 & ~3.65 & NV & $-$ & $-$  & $-$  & $-$                     \\
0414190501 & 3.19$\pm$0.10 & 31.9 &  ~V & 2.50$\pm$0.27 & 9.26 &  ~V & 3.04$\pm$0.09 & 34.26 & ~V & 1.17$\pm$0.32 & 1.01$\pm$0.28 & 1.09$\pm$0.32 & 0.71$\pm$0.26 \\
0414190601 & 0.53$\pm$0.22 & 2.41 &  NV & 1.35$\pm$0.37 & 3.65 &  NV & 0.73$\pm$0.14 & ~5.13 & ~V & 1.58$\pm$0.51 & 1.36$\pm$0.44 & $-$ & $-$ \\
0414190701 & 0.53$\pm$0.22 & 2.41 &  NV & 0.91$\pm$0.63 & 1.44 &  NV & 0.83$\pm$0.15 & ~5.46 & ~V & 1.29$\pm$0.39 & 1.11$\pm$0.34 & $-$ & $-$     \\
0414190801 & 1.71$\pm$0.12 & 14.25&  ~V & 3.84$\pm$0.25 & 15.36&  ~V & 2.07$\pm$0.10 & 20.95 & ~V & 1.49$\pm$0.55 & 1.29$\pm$0.48 & 1.11$\pm$0.34 & 0.50$\pm$0.15   \\
0414191001 & 0.30$\pm$0.66 & 0.45 &  NV & 0.75$\pm$1.16 & 0.65 &  NV & 0.22$\pm$0.73 & ~0.30 & NV & $-$ & $-$  & $-$ & $-$          \\
0414191101 & 1.04$\pm$0.15 & 6.93 &  ~V & 1.40$\pm$0.46 & 3.04 &  NV & 1.08$\pm$0.13 & ~8.52 & ~V & 1.13$\pm$0.37 & 0.98$\pm$0.32 & 1.03$\pm$0.34 &  $-$   \\
0414191201 & 0.49$\pm$0.18 & 2.72 &  NV & 0.70$\pm$0.45 & 1.56 &  NV & 0.53$\pm$0.14 & ~3.87 & NV & $-$ & $-$  & $-$ & $-$     \\
0414191301 & 0.29$\pm$0.53 & 0.55 &  NV & 2.31$\pm$0.31 & 7.45 &  ~V & 0.68$\pm$0.20 & ~3.47 & NV & $-$ & $-$  & $-$ & 0.48$\pm$0.16    \\
0414191401 & 0.95$\pm$0.23 & 4.13 &  NV & 1.05$\pm$0.69 & 1.52 &  NV & 0.85$\pm$0.20 & ~4.18 & NV & $-$ & $-$  & $-$ & $-$ \\
0810820101 & 0.48$\pm$0.43 & 1.12 &  NV & 0.75$\pm$1.04 & 0.72 &  NV & 0.36$\pm$0.46 & ~0.78 & NV       & $-$ & $-$  & $-$ & $-$  \\
0810821501 & 1.79$\pm$0.14 & 12.78&  ~V & 2.09$\pm$0.36 & 5.81 &  ~V & 1.87$\pm$0.12 & 16.09 & ~V & 1.00$\pm$0.33 & 0.86$\pm$0.28 & 0.84$\pm$0.26 & 0.51$\pm$0.18 \\
0810821601 & 0.83$\pm$0.35 & 2.37 &  NV & 1.04$\pm$0.99 & 1.05 &  NV & 0.72$\pm$0.31 & ~2.29 & NV & $-$ & $-$ & $-$ & $-$ \\\hline
\end{tabular}}

F$_{var}$ = fractional variance\\
$^{a}$ = Significance (Sig) = F$_{var}$/(F$_{var}$)$_{err}$ \\
$^{b}$ = Variable (Var) = V = variable, NV = non-variable \\
$^{c}$ = ${|\tau|}$= observed halving/doubling time-scale in ks. \\
$^{d}$ = $|\tau|_{corr}$ = Redshift corrected halving/doubling time-scale:$|\tau|_{cor}$ = $|\tau |$/(1 + z)

\end{table*}

\section{Analysis Techniques}
In this section, we introduce the analysis methods used to evaluate various quantities characterizing this X-ray data from the blazar 3C 273. The results we obtained are reported in section 4.

\subsection{Excess and Fractional Variance}
To measure the strength of blazar variability in X-ray bands the commonly used parameters are excess variance $\sigma_{XS}^{2}$, and fractional rms variability amplitude $F_{var}$ \citep[e.g.,][]{Edelson2002}. Excess variance provides the intrinsic variance of the source by removing the variance due to measurement errors in each individual flux measurement and fractional variance gives the mean variability amplitude with respect to the mean flux of the source. When the LC contains the total number $n$ of flux measurements $x_{i}$ at times $t_{i}$ with corresponding errors in measurements $\sigma_{err,i}$, and mean ${\bar x}$, then the excess variance is given by
\begin{equation}
\sigma_{XS}^{2}=S^{2}-\bar \sigma_{err}^{2} ~,
\end{equation}

\noindent
where $\bar \sigma_{err}^{2}$ is the mean square error and $S^{2}$ is the sample variance of the LC \citep[for details see][]{Edelson2002,2021MNRAS.506...1198D}.\\ 

\noindent
The fractional variance is given by
\begin{equation}
F_{var}=\sqrt{\frac{S^{2}-\bar \sigma_{err}^{2}}{\bar x^{2}}}
\end{equation} 

\noindent
and error in fractional variance is given by \citep[e.g.][]{2003MNRAS.345.1271V}
\begin{equation}
(F_{var})_{err}=\sqrt{\left[\sqrt \frac{1}{2n}\frac{\bar \sigma_{err}^{2}}{F_{var}\bar x^{2}}\right]^2 +\left[\sqrt{\frac{\bar \sigma _{err}^{2}}{n}}\frac{1}{\bar x}\right]^{2}}
\end{equation}
The values of excess and fractional variances for the soft, hard, and total X-ray bands are given in Table 2.

\subsection{Variability Timescale}

We use the method for estimating variability timescales as described in \citep{2018A&A...619A..93B,2019ApJ...884..125Z}. As explained in \citet{1974ApJ...193...43B}, the timescale of variability of the flux is given by
\begin{equation}
\tau_{var}=\left |\frac{\Delta t}{\Delta \ln\frac{F_1}{F_2}} \right |,
\end{equation}
where $\Delta$t is the time interval between the flux measurements $F_1$ and $F_2$ with $F_1> F_2$. The error in the $\tau_{var}$ is given by the formula \citet{2019ApJ...884..125Z}
\begin{equation}
\Delta \tau_{var}\approxeq \sqrt{\frac{F_{1}^{2}\Delta F_{2}^{2}+F_{2}^{2}\Delta F_{1}^{2}}{F_{1}^{2}F_{2}^{2}(ln[F_{1}/F_{2}])^{4}}} \quad \Delta t
\end{equation}
Here $F_{1}$ and $F_{2}$ are the count rates used to estimate the shortest variability timescales and $\Delta F_{1}$ and $\Delta F_{2}$ are their corresponding uncertainities. As described in \citet{2008ApJ...672...40H}, for a given observation, $F_1$ and $F_2$  are chosen in such a way that $|F_i-F_j|> \sigma_{F_i}+ \sigma_{F_j}$ where $\sigma_F$ is the error in flux measurment. The minima of all such pairs $\tau=min \{ \tau_{ij} \}$, where $i=1,...N-1,j=i+1....N$ and $N$ is the number of flux values, is the value of the shortest variability timescale. The variability timescale values are given in Table 2. 

\subsection{Power Spectral Density}
The power spectral density (PSD) provides the distribution of variability power as a function of temporal frequency and typically displays a red-noise character at lower frequencies, transitioning to white-noise at higher frequencies where measurement errors dominate. \\
\\
We compute the PSD using the \emph{periodogram} routine in {\it python} where the normalization used to plot the periodogram is (rms/mean)$^{2}$ Hz$^{-1}$. Then, we fit the periodogram data using maximum likelihood estimation. As discussed in \citet{2010MNRAS.402..307V}, the best fitting model parameter $\theta$ is evaluated by maximizing the likelihood function. This is done by minimizing the following fit statistic:
\begin{equation}
S=2\sum_{j=1}^{N/2}\frac{I_{j}}{P_{j}}+\ln{P_j}
\end{equation}
where $S$ is the twice the negative of the  log-likelihood and $I_j$ and $P_j$ are the observed periodogram and model spectral density at Fourier frequency $\nu_j$, respectively. Confidence intervals on each model parameter correspond to 68.3$\%$ ($1\sigma$). A significant QPO may be present whenever a peak rises at least by $3\sigma$ ($99.73\%$) above the red noise level of the PSD. In the literature, a power-law 
model has been used to fit PSDs of the LCs in the X-ray and other bands \citep[e.g.][]{Gonzales2012,Mohan2015}. The power-law form we used is given by
\begin{equation}
P(\nu)=N \nu^{-\alpha}+C,
\end{equation}
\noindent
where $N$ is the power-law normalization, $\alpha$ is the spectral index, and $C$ is the additive constant accounting for Poisson noise. We take the value of $C$ as a free parameter that is used for fitting the data. The  power-law PSD fits are discussed in Section 4.4. 

\subsection{Discrete Correlation Function}

We use the standard formula for the unbinned discrete correlation function (UDCF) between the hard and soft counts \citep{1988ApJ...333..646E,1992ApJ...386..473H}
 \begin{equation}
UDCF_{ij}=\frac{(x_i-\bar x)(y_i-\bar y)}{\sqrt{\sigma_x^2 \sigma_y^2}}.
\end{equation}
Here $x_i$ and $y_i$ are hard and soft data points, and $\bar x$, $\bar y$, $\sigma_{x}$, $\sigma_{y}$ are their means and standard deviations. respectively. Each of the values is associated with a pair-wise lag $t_{ij}=t_j-t_i$. After calculating the UDCF, the DCF at a time-lag $\tau$, where  $\tau - \frac {\delta \tau}{2} \leq \delta t_{ij} \leq \tau+ \frac {\delta \tau}{2}$ is given by
\begin{equation}
DCF \left (\tau \right )=\frac{1}{M} \sum UDCF_{ij}
\end{equation}
with a bin value $M$ selected as 10.
\noindent 
The error is defined as
\begin{equation}
\sigma_{DCF} \left (\tau \right )=\frac{\sqrt{\sum (UDCF_{ij}-DCF(\tau))}}{M-1}
\end{equation}

\noindent
In general, a positive DCF peak means the two data sets are correlated, negative means they are anti-correlated and no DCF peak or DCF = 0 means no correlation exists between the two data sets. The correlation results are discussed in Section 4.5.

\subsection {Hardness Ratio}
The hardness ratio (HR) is defined as
\begin{equation}
HR=\frac {(H-S)}{(H+S)}
\end {equation}
where $H$ and $S$ are, respectively, the net count rates in the hard and soft bands. The error on this quantity, $\sigma_{HR}$ is calculated as \citep[e.g.][]{2021ApJ...909..103Z}
\begin{equation}
\sigma_{HR}=\frac{2}{(H+S)^2}\sqrt {(H^2\sigma_{S}^{2}+S^2\sigma_{H}^{2})}.
\end {equation}
HR provides a simple and model independent way to study the spectral variability of a source. The HR plot for the Obs IDs we have examined are shown in Fig.\ 3 and in the Supplemental Material.

\subsection{Duty Cycle}
The duty cycle (DC) is the fraction of time when an object displays variability. We have estimated the DC of X-ray variability in 3C 273  using the approach given in \citet{Romero1999} and commonly followed thereafter. For calculations of the DC, only those LCs that were continuously monitored for at least 6 hours were considered. The formula is 
\begin{equation}
DC=100\frac{\Sigma_{i=1}^{n} N_{i}(1/\Delta t_{i})}{\Sigma_{i=1}^{n}(1/\Delta t_{i})} \% ,
\end{equation}

\noindent
where $\delta t_{i} = \delta t_{i,obs}(1 + z)^{-1}$ is the redshift corrected GTI of the source observed having the $i^{th}$  Ob ID, and $N_{i}$ takes the value 1 whenever IDV is detected, and 0 when it is not. For a particular Obs ID, if the  $F_{var}$ is greater than 5 times the value of $(F_{var})_{err}$, then we consider the LC to show genuine IDV and the $N$ value is taken as 1, as noted in Table 2. 

\begin{figure}
\centering
\vspace{-1.4cm}
\includegraphics[scale=0.4]{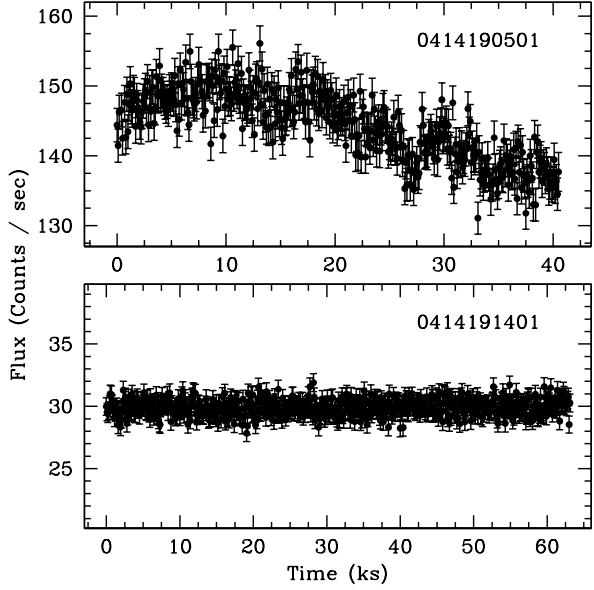}
\vspace{-1.0in}
\caption{A sample {\it XMM-Newton} observation ID 0414190501 (variable) and observation ID 0414191401 (non variable) 
LCs of the blazar 3C 273 in the total (0.2 -- 10 keV) energy band, labelled with its observation ID. The LCs 
for all observations appear in online supplemental material.}
\end{figure}

\begin{figure}
\centering
\vspace{-1.0cm}
\includegraphics[scale=0.4]{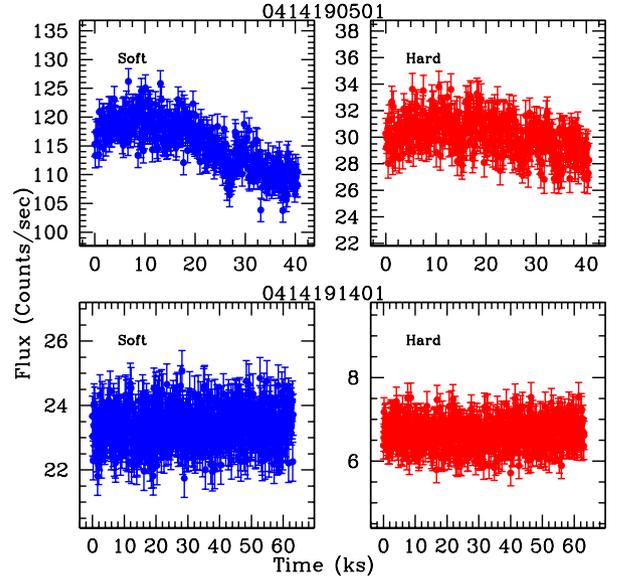}
\vspace{-1.0in}
\caption{Soft energy (0.2 - 2 keV, upper left plot) and hard energy (2 - 10 keV, upper right plot) LCs for the blazar 3C 273 for a sample {\it XMM-Newton} observation ID 0414190501 (variable). The similar plots are presented in the lower panels for observation ID 0414191401 (non variable) LCs. The LCs for all observations appear in online supplemental material.}
\end{figure}

\section{Results}
By applying the various analysis methods described in Section 3 to the {\it XMM-Newton} data listed in Table 1, we obtain the following results.

\subsection{Intraday Flux Variability} 
We generated X-ray LCs of individual observation IDs with these 23 observations using three {\it XMM-Newton} EPIC-pn energy bands (soft, hard, and total). Exemplary LCs of  observation IDs 0414190501 (variable) (top panel) and 0414191401 (non variable) (bottom panel) in the total energy (0.2 -- 10 keV) band are shown in Fig.\ 1; the corresponding  soft (0.2 -- 2.0 keV) and hard (2 -- 10 keV) bands are shown in Fig.\ 2.    
All 23 LCs are displayed in online supplemental material. \\
\\
To investigate the flux variability of the blazar 3C 273 on IDV timescales and estimate the variability amplitudes, we have followed the excess variance method explained in Section 3.1 and the obtained results are reported in Table 2.  We consider a LC is variable when F$_{var} >$ 5 $\times$ (F$_{var})_{err}$. Using this method, out of 23 LCs in each soft (0.2 -- 2 keV), hard (2 -- 10 keV), and total (0.2 -- 10 keV) energies, we found that 8, 5, and 9 LCs show IDV, respectively. Both well defined increase and decrease in flux within a single observation and rising or falling trends spanning the observation are included with this definition of IDV.  We estimated F$_{var}$ and its error for the soft, hard and total bands for all individual Obs IDs and these are reported in Table 2. We also report there the variability time scales, as defined in section 3.2, for all the LCs showing IDV in soft, hard and total energy bands.   

\begin{figure}
\centering
\vspace{-1.0cm}
\includegraphics[scale=0.4]{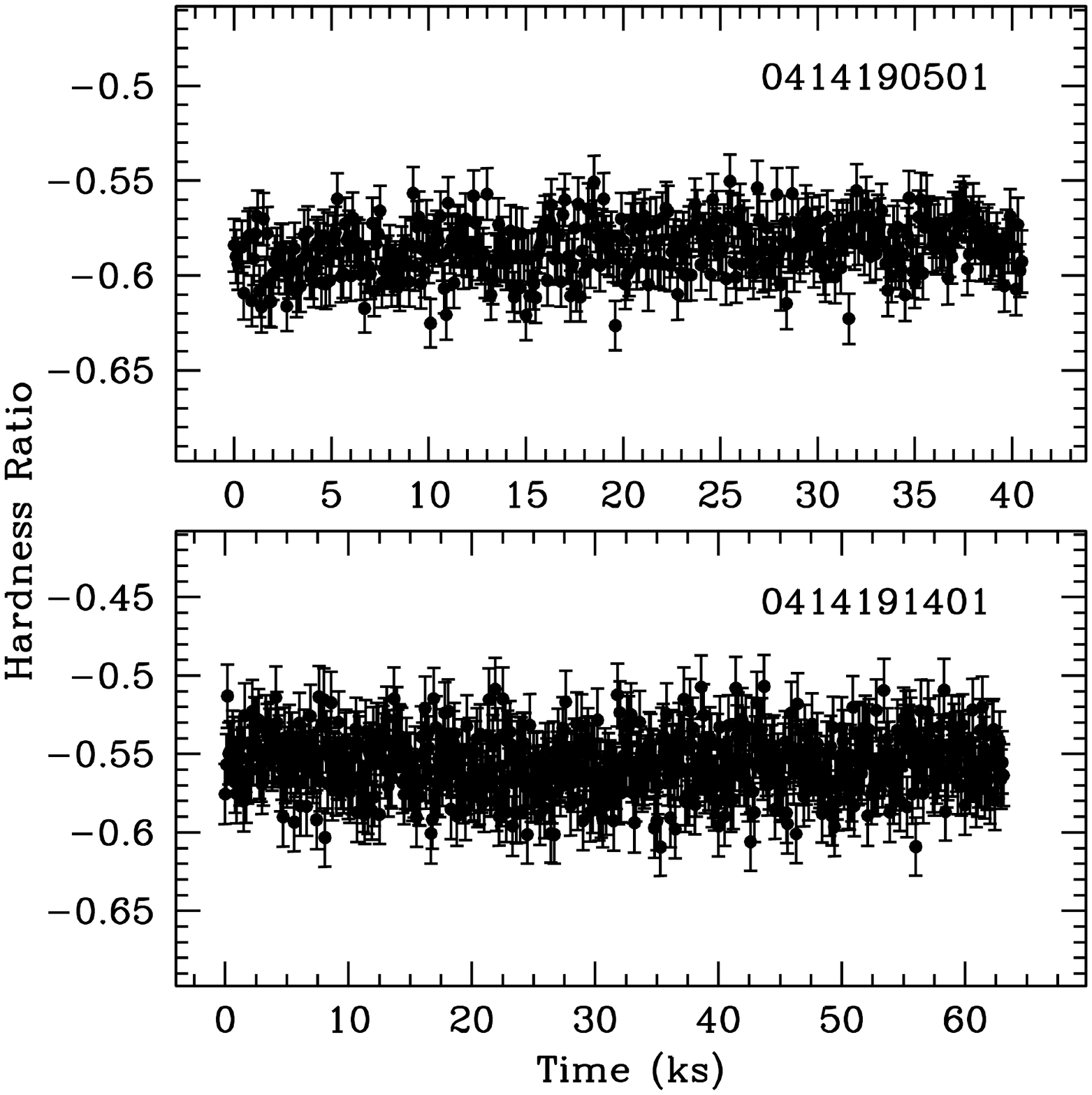}
\vspace{-1.0in}
\caption{Hardness ratio plot for soft energy (0.2 - 2 keV) and hard energy (2 - 10 keV) for LCs for the blazar 3C 273 for a sample {\it XMM-Newton} observation ID 0414190501 (variable LCs) and observation ID 0414191401 (non variable LCs).  The hardness ratio plots for all observations appear in the online supplement material.}
\end{figure}

\subsection{Intraday Spectral Variability}
The hardness ratio (HR) is the simplest and a valuable model independent tool to represent the spectral variations of X-ray emission. The HR with respect to time for exemplary observation IDs 0414190501 (for variable LCs) (top panel) and 0414191401 (for non variable LCs (bottom panel) are plotted in Fig.\ 3. All 23 HRs for individual Obs IDs are displayed in online supplemental material. We found no significant changes in HR with respect to time for any of these 23 LCs, which is not surprising since these LCs never showed large amplitudes of variability.

\subsection{Duty cycle}
We estimated the X-ray variability DC of 3C 273 using the method mentioned in Section 3.6. By considering $F_{var}$ in comparison to its errors to define variability we took the value of $N_{i} = 1$ if variable, or  0 if not. In the soft (0.2 -- 2 keV) and hard (2 -- 10 keV) energies, 6 out of 23 and 5 out of 23 LCs, respectively, exhibited variability. But in the total X-ray energy range (0.2 -- 10 keV) where the count rates are higher, 9 out of 23 observations were variable. Using this method we found that the X-ray DC of 3C 273 during 2000 -- 2021 to be $\sim$26 per cent, $\sim$22 per cent and $\sim$39 per cent in the soft, hard and total energy bands, respectively. However, only 4 out of these 23 Obs IDs have shown clear variations in all three energy bands.   

\begin{table*}
{\bf Table 3.} Parameters of the power law fits to the PSDs in soft, hard and total bands of 3C 273.
\centering
\noindent
\scalebox{1.2}{
\begin{tabular}{l c c c c c c} \hline \hline
 ObsID  & \multicolumn{2}{c} {Soft (0.2-2 keV)} & \multicolumn{2}{c} {Hard (2-10 keV)} & \multicolumn{2}{c} {Total (0.2-10 keV)} \\\cline{2-7}
       &                $\log_{10} N$    & $ \alpha $     &     $\log_{10} N$    & $ \alpha $  & $\log_{10} N$    & $ \alpha $  \\ \hline%
0126700301 & $-$ & $-$ & NV & NV & NV & NV \\
0126700601 & NV & NV & NV & NV & NV & NV \\
0126700701 & NV & NV & NV & NV & NV & NV \\
0126700801 & $-$6.33$\pm$0.29 & 1.43$\pm$0.06 & NV & NV & $-$7.64$\pm$0.44 & 1.71$\pm$0.09   \\
0136550101 & $-$12.99$\pm$0.55 & 2.77$\pm$0.11 & NV & NV & $-$12.38$\pm$0.63 & 2.61$\pm$0.13   \\
0159960101 & NV & NV & NV & NV &  NV &  NV \\
0136550801 & NV & NV & NV & NV &  NV &  NV \\
0136551001 & NV & NV & NV & NV &  NV &  NV \\
0414190101 & $-$10.88$\pm$0.40 & 2.38$\pm$0.08 & $-$9.22$\pm$0.40 & 2.13$\pm$0.08 & $-$8.81$\pm$0.42 & 1.96$\pm$0.09   \\
0414190301 & NV & NV & NV & NV &  NV &  NV \\
0414190401 & NV & NV & NV & NV &  NV &  NV \\
0414190501 & $-$9.19$\pm$0.61 & 2.29$\pm$0.13 & $-$6.09$\pm$0.61 & 1.54$\pm$0.13 & $-$7.67$\pm$0.93 & 1.93$\pm$0.20   \\
0414190601 & NV & NV & NV & NV & $-$10.92$\pm$1.18 & 2.44$\pm$0.26   \\
0414190701 & NV & NV  & NV & NV & $-$8.13$\pm$0.88 & 1.80$\pm$0.20   \\
0414190801 & $-$12.00$\pm$0.85 & 2.79$\pm$0.18 & $-$11.54$\pm$0.85 & 2.85$\pm$0.10 & $-$11.37$\pm$0.61 & 2.69$\pm$0.13   \\
0414191001 & NV & NV & NV & NV & NV & NV \\
0414191101 & $-$ & $-$ & NV & NV & $-$6.77$\pm$0.26 & 1.52$\pm$0.05 \\
0414191201 & NV & NV & NV & NV & NV & NV \\
0414191301 & NV & NV & $-$13.66$\pm$1.06 & 3.12$\pm$0.22 & NV & NV   \\
0414191401 & NV & NV & NV & NV &  NV &  NV \\
0810820101 & NV & NV & NV & NV &  NV &  NV \\
0810821501 & $-$12.57$\pm$0.53 & 2.84$\pm$0.11 & $-$10.52$\pm$0.55 & 2.48$\pm$0.11 & $-$11.67$\pm$0.27 & 2.67$\pm$0.06   \\
0810821601 & NV & NV & NV & NV & NV & NV \\\hline
\end{tabular}} \\
Note: $N$ is the normalization and $\alpha$ is the spectral index \\
$-$ indicates the variations were too small to compute a PSD \\
NV indicates that the observation is non-variable
\vspace{1cm}
\end{table*}

\begin{figure}
\centering
\includegraphics[scale=0.45]{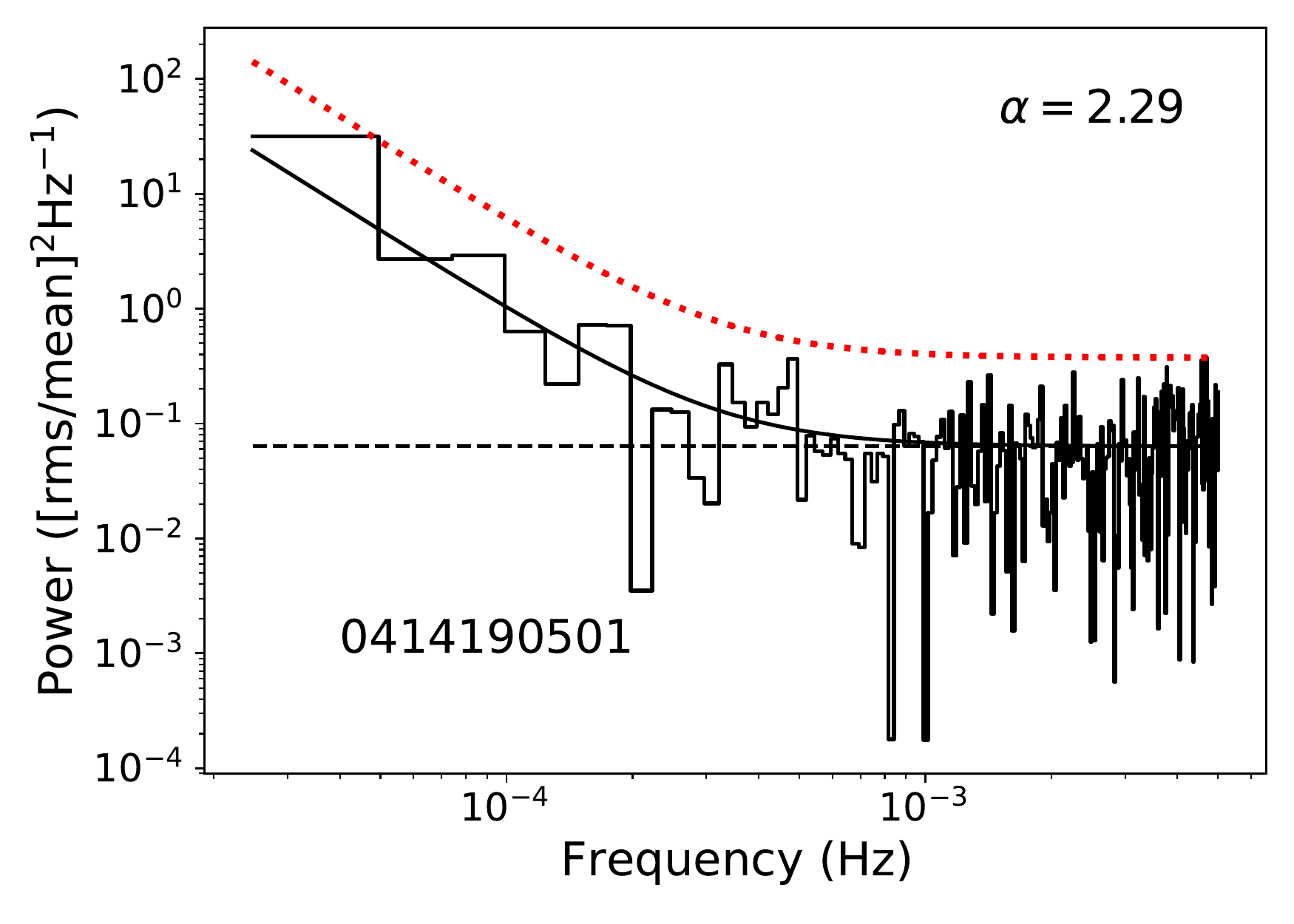}
\vspace{-0.2cm}
\caption{Power spectral density plot of the LC for a sample {\it XMM-Newton} observation ID 0414190501 (variable LC) in the soft energy band. The observational ID and power spectral index are given in the plot. The continuous line shows the power-law fit to the red noise and the red dotted line shows a level 3$\sigma$ above the red noise level. PSD plots for all observations appear in online supplement material.} 
\end{figure}

\begin{figure}
\centering
\includegraphics[scale=0.45]{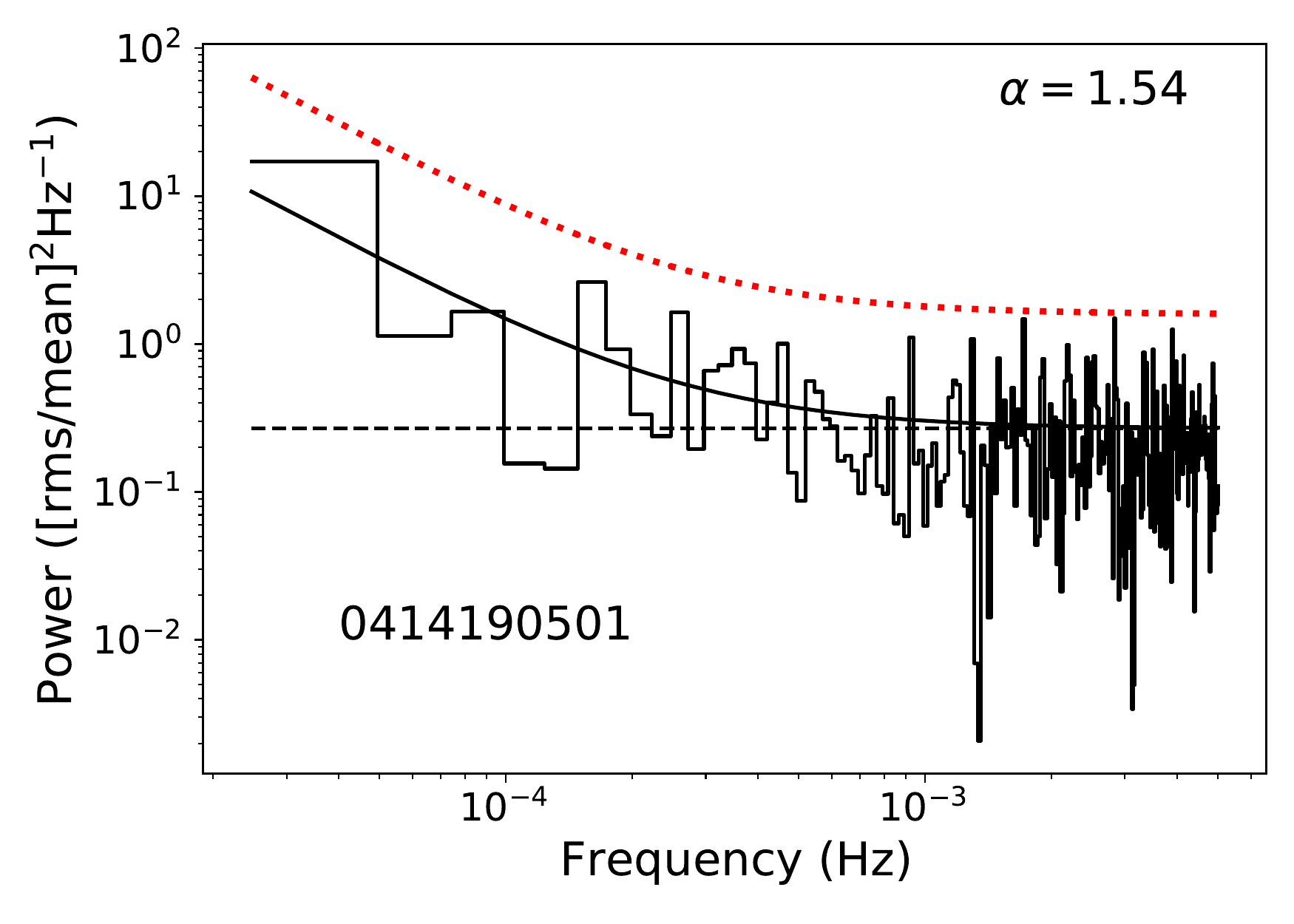}
\vspace{-0.2cm}
\caption{Power spectral density plot of the LC for a sample {\it XMM-Newton} observation ID 0414190501 (variable LC) in the hard energy band. The observational ID and power spectral index are given in the plot. The continuous line shows the power-law fit to the red noise and the red dotted line shows a level 3$\sigma$ above the red noise level. PSD plots for all observations appear in online supplement material.}
\end{figure}

\begin{figure}
\centering
\includegraphics[scale=0.45]{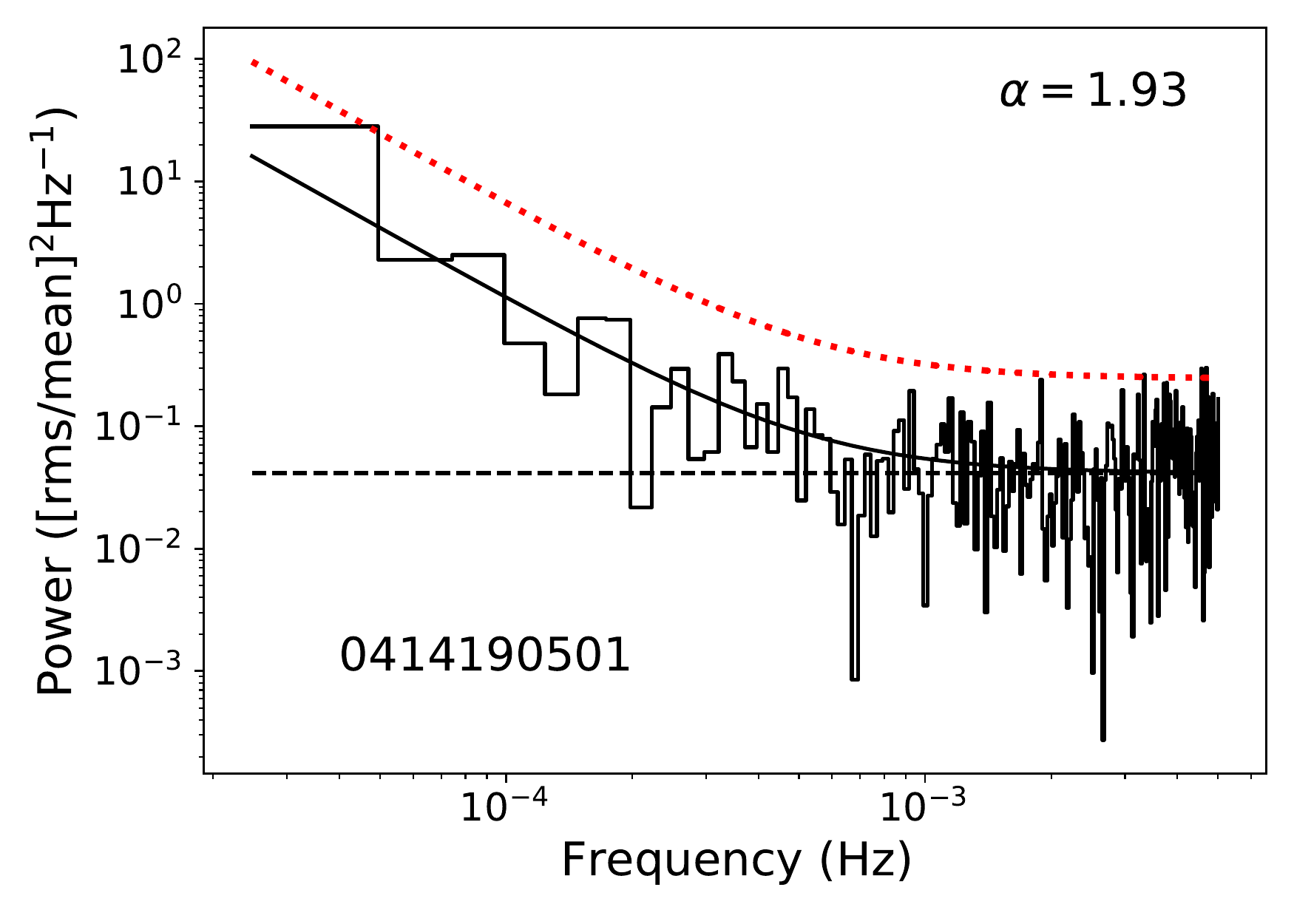}
\vspace{-0.2cm}
\caption{Power spectral density plot of the LC for a sample {\it XMM-Newton} observation ID 0414190501 (variable LC) in the total energy band. The observational ID and power spectral index are given in the plot. The continuous line shows power-law fitting of the red noise and the red dotted line shows a level 3$\sigma$ above the red noise level. PSD plots for all observations appear in online supplement material.}
\end{figure}

\subsection{Intraday Power Spectral Density Analysis}

In a major analysis of a large sample of X-ray LCs of various subclasses of AGNs, it was confirmed that the power spectral densities (PSDs) are red-noise dominated that decreases steeply over a range of frequencies as a power-law P($\nu) \sim \nu^{-\alpha}$ (where $\nu$ is temporal frequency), typically with spectral index $\alpha \approx$ 2 \citep{Gonzales2012}. Below some particular frequency $\nu_{b}$ (bend frequency), the PSDs flatten, and these $\nu_{b}$ values scale approximately inversely with the SMBH mass for AGN \citep{Gonzales2012}. Here we characterize the PSDs slopes of the blazar 3C 273 for all nights on which it showed IDV in soft, hard, and total energies  taken by {\it XMM-Newton} over the course of its whole operational period. The resulting  fit parameters are given in Table 3, where the slopes and normalization for the power-law red-noise portion of the PSD are given before the flattening to white-noise at the highest frequencies.  These power-law fits, to the PSDs observation ID 0414190501 are shown for soft, hard, and total bands in Figures 4 -- 6, respectively. The PSDs from all the Obs IDs with detected variability during the course of an observation are displayed in the online supplemental material.  
We see no evidence for a QPO in any of these PSDs. \\
\\
In Fig.\ 7 we have plotted the PSD indices, $\alpha$, against the fluxes for the 4 Obs IDs in which soft, hard and total X-ray energies are all variable. These points appear show a  trend in the sense that the brighter sources have shallower PSD slopes. We fitted the straight line in PSD index $\alpha$ versus flux plots in soft, hard and total energies for these 4 obs IDs and in Table 4 where we provide slope, correlation coefficients and probabilities for the null hypothesis ($p$-value).  However, there are too few data points for any  conclusions to be drawn about any trend. 

\begin{table}
{\bf Table 4.} Spectral index versus flux correlation parameters for soft, hard and total X-ray energies.
\centering
\noindent
\scalebox{1.2}{
\begin{tabular}{l c c c} \hline \hline
Energy   &  Slope               & Correlation & p-value \\ 
band     &                      & coefficient &         \\\hline
Soft     &  $-$0.005$\pm$0.004  & 0.451       & 0.328    \\
Hard     &  $-$0.051$\pm$0.020  & 0.766       & 0.125    \\
Total    &  $-$0.009$\pm$0.006  & 0.498       & 0.294   \\\hline
\end{tabular}} \\
\end{table}

\begin{figure}
\centering
\vspace{-0.8cm}
\includegraphics[scale=0.4]{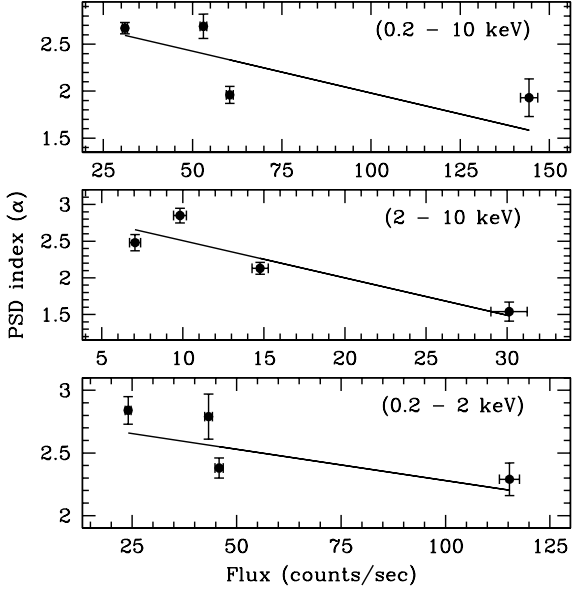}
\vspace{-2.5cm}
\caption{PSD spectral index versus flux plots of the 4 Obs ids which have shown IDV in all soft, hard, and total X-ray energies are plotted with  filled circles and straight line fits to possible trends.}  
\end{figure}

\subsection{Intraday Cross-correlated Variability}
The relationship between X-ray fluxes in the soft and hard bands is shown in Fig.\ 8. The plot shows a tight correlation with correlation coefficient 0.953 and probabilities for the null hypothesis ($p$-value) of $2.3 \times 10^{-12}$. This reflects the fact that the HR values changed little between the observations as well as during each one. Such tight correlation between the soft and hard X-ray bands supports the hypotheses that these photons originate from the same regions and are emitted from the same population of leptons. Even stronger support for that likely situation could come from tight cross-correlations between the soft and hard X-ray bands in each individual observation at a very small temporal lag. To test this we performed DCF analyses between these two bands; however, several of the LCs do not show any genuine variability, and when variability is detected, the variability amplitude is usually small. So this analysis only produced poor and irregular DCF plots which are unlikely to indicate real correlations. Similar results were also reported for some of the common {\it XMM-Newton} observation IDs of 3C 273 reported previously \citep{Kalita2015}. 

\begin{figure}
\centering
\vspace{-1.0cm}
\includegraphics[scale=0.4]{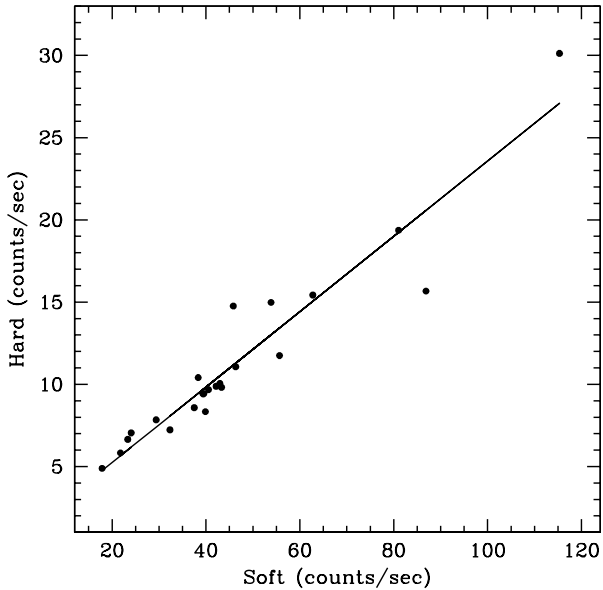}
\vspace{-2.5cm}
\caption{Hard energy flux versus soft energy flux plot for all the observations of the blazar 3C 273.} 
\end{figure}

\subsection{Long-term Flux and Spectral Variability}
These pointed observations of the blazar 3C 273 were carried out by {\it XMM-Newton} on many occasions over more than two decades. These observations provide us an excellent opportunity to study its X-ray flux and spectral variability on LTV timescales. The overall variation in the total X-ray flux with respect to time is shown in the top panel of Fig.\ 9. On visual inspection we notice that there appears to be a weak trend of decreasing flux with respect to time. There is however, one particularly high X-ray flux point, marked by a red square, that can be considered as an outlier. We computed a least-square fit to these long term data of flux against time, excluding that outlier, which yields a slope of 
$-$1.4 counts s$^{-1}$ yr$^{-1}$ with a correlation coefficient of the fit of $-$0.487 and a corresponding null hypothesis of 0.0216. This least-square fit result allows us to say that there is indeed a weak decreasing trend in the flux with time. \\
\\
The HR (overall spectral change) with time over the entire duration of these observations is presented in the bottom panel of Fig.\ 9. On visual inspection we notice that there is weak trend of increasing HR with respect to time. One high value HR point is considered as an outlier and marked by red square.  The least-square fit to these long term data of HR versus time (excluding the outlier) results in a slope of 
3 $\times$ 10$^{-3}$ yr$^{-1}$ which indicates a decline of 0.06 over $\sim$ 21 years, with correlation coefficient of the fit of 0.631 and its corresponding null hypothesis of 0.0016. With these least-square fit results, we can say that there is a  weak increasing trend in the HR with time on this long timescale. Together, the panels of Fig.\ 8 indicate an anti-correlation in flux and HR, or a harder-when-brighter trend.  This confirms the result found in \citep{Kalita2015} for the data taken during 2000 -- 2015. \\
\\
In Fig.\ 10, we present plots of HR against flux of the blazar 3C 273 for six temporal intervals. Spectral evolution of 3C 273 during $\approx$ 21 years of observation can be interpreted in terms of the changing relative importances of particle acceleration and synchrotron cooling processes in X-ray emitting regions \citep{Kalita2015}. In Fig.\ 10, we note indications of anticlockwise loops (hard lag) in epochs 2, 3, 4 and partially in epochs 5 and 6. The anticlockwise loops represent that during these epochs of observations the mechanism dominating emission is particle acceleration  \citep[e.g.,][and references therein]{2002ApJ...572..762Z}. In epoch 1 and arguably in parts of epochs 5 and 6  clockwise loops (soft lags)  are seen in Fig.\ 10; this can be understood as time when the  synchrotron cooling mechanism is dominant. Similar results were found for a subset of these observations of 3C 273 from {\it XMM-Newton} \citep{Kalita2015}, as well as for Mrk 421 from {\it Chandra} \citep{2018MNRAS.480.4873A}, and PKS 2155-304 from {\it Suzaku} \citep{2021ApJ...909..103Z}.   

\begin{figure}

\centering
\includegraphics[scale=0.4]{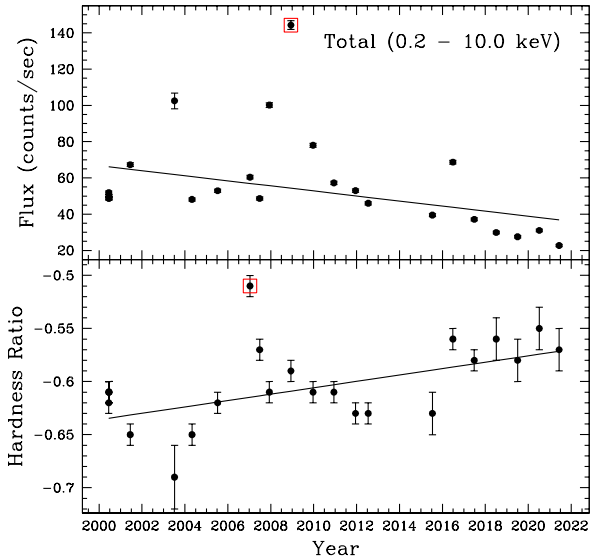}
\vspace*{-2.4cm}\caption{Long term X-ray flux variability (top panel) and spectral variability (bottom panel) of 3C 273.}
\end{figure}

\begin{figure}
\centering
\includegraphics[scale=0.4]{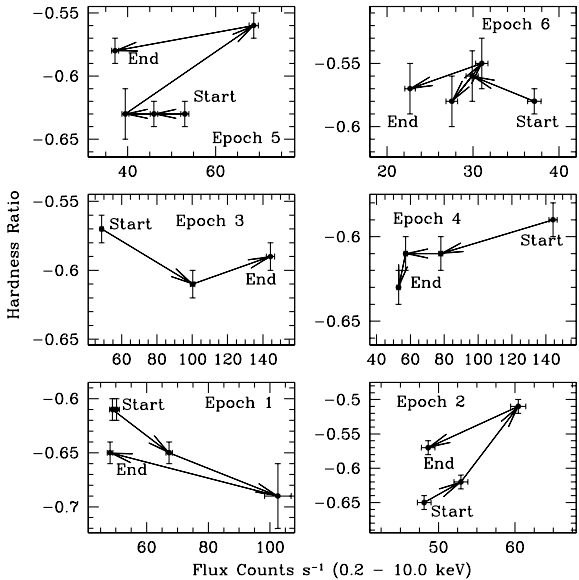}
\vspace*{-1.8cm}\caption{Spectral variations of 3C 273 in different epochs with start and end points marking the loop directions. Each epoch corresponds to the time interval during which the data were considered from Epoch 1 to Epoch 6: Epoch 1: 2000 June 13 to 2004 June 30; Epoch 2: 2004 June 30 to 2007 June 25; Epoch 3: 2007 June 25 to 2008 December 9; Epoch 4: 2008 December 9 to 2011 December 12; Epoch 5: 2011 December 12 to 2017 June 26; and Epoch 6: 2017 June 26 to 2021 June 9.}
\end{figure}

\section{Discussion}
We have studied the 23 longest archival pointed observations of the FSRQ blazar 3C 273 taken during 2000 to 2021 by the EPIC-pn instrument on board  {\it XMM-Newton}. These observations were carried out with GTIs from 18.0 to 88.6 ks. We aimed to study flux and spectral variability, cross-correlated variability, and PSDs analyis on IDV timescales for this blazar. \\
\\
In a series of ten {\it XMM-Newton} observations in which four are in common with the present study, the soft excess component was found to vary, and could be well fitted by multiple black body components, with temperatures ranging between $\sim$40 and $\sim$330 eV, together with a power-law \citep{2004MNRAS.349...57P}. By using quasi-simultaneous {\it INTEGRAL} and {\it XMM-Newton} monitoring of the blazar 3C 273 in 2003 -- 2005 the 0.2 -- 100 keV spectrum of the source was well fitted by a combination of a soft cut-off power-law and a hard power-law and the source reached its historically softest state in the hard X-ray domain with a photon index $\Gamma =$ 1.82$\pm$0.01 \citep{2007A&A...465..147C}. There were 3 common Obs IDs of {\it XMM-Newton} in the present study and that of \citet{2007A&A...465..147C}.  \\
\\
Multi-band X-ray, UV and optical observations of 3C 273 including data from {\it XMM-Newton} were used to study flux, spectral and cross-correlated variabilities on diverse timescales. Long-term observations of 3C 273 found that IDV X-ray LCs showed small amplitude variability, along with large amplitude variability in optical to X-ray energies on longer timescales, an anti-correlation in UV and X-ray emission in a low flux state, and a harder-when-brighter trend for the X-ray spectrum \citep{Kalita2015,2017MNRAS.469.3824K}. There are 16  Obs IDs studied in the present paper that are also considered in those papers \citep{Kalita2015,2017MNRAS.469.3824K}. \citet{Gonzales2012} have studied PSD properties of 104 AGN including 3C 273 using {\it XMM-Newton} observations, including 6 common Obs IDs studied in the present paper and in that one. The PSD results of \citet{Gonzales2012} and in the present study for common Obs IDs are consistent. \\
\\
Therefore, the modest X-ray variability seen in 3C 273 seen on these timescales might arise from the jet or accretion disc coronal emission, or a combination of both.
Characteristics of such variations, particularly their PSDs, might be used to distinguish between those possibilities.  While there have now been a substantial number of jet and disc simulations, including some very sophisticated general relativistic magnetodynamical computations including radiation transport that include disc, corona, and the innermost portion of a jet \citep[e.g.,][]{2021Sci...373..789B} only a very small subset of them carry these simulations far enough to yield light curves and then compute their power-spectra. As summarized in \citet{2019ApJ...877..151W} accretion disc models that do so typically produce PSD slopes in the range $1.3 \lesssim \alpha \lesssim 2.1$, though it is  possible to see somewhat steeper PSDs for some geometries.  On the other hand, jet based models more typically yield somewhat steeper PSD slopes  $1. 7 \lesssim \alpha \lesssim  2.9$, although these simulations have typically probed timescales from weeks to years.  The observed PSD slopes found here ($1.43 \le \alpha \le 3.12$)  for the X-ray emission of 3C 273 span both ranges, though the majority of them  seem to be more consistent with jet-based simulations.  However, given the limited range of the PSDs found in this work and the still small number of  computational papers that provide PSDs, we cannot claim more than a tentative hint favouring fluctuations originating in jets. 

\section{Conclusions}
We studied 23 observations of the FSRQ  3C 273 from the public archive {\it XMM-Newton}. These EPIC-pn  light curves  for the blazar  were taken during 2000 -- 2021. We searched for intraday variability and its timescales, HRs, time lags between soft and hard energies, and also carried out PSD analyses to characterize the IDV and to search for any possible QPO present. Our conclusions are summarized as follows: \\
\\
$\bullet$ The source showed fractional variabilities of small amplitude 0.71$\pm$0.10 per cent to 3.04$\pm$0.09 per cent in 9 light curves, while 14 light curves do not show any significant variability in the total X-ray energy band. This indicates a duty cycle of $\sim$ 40 per cent in the total energy band. No strong flare is seen on any of the observation ID light curve and the typical IDV timescale is found to be $\sim$ 1ks.  \\
\\
$\bullet$ There is no significant change in spectral variation, as measured by the hardness ratio, seen during any of the observations. \\
\\
$\bullet$ The relationship between X-ray fluxes in the soft and hard bands shows a tight correlation with correlation coefficient 0.953 and probability for the null hypothesis (p-value) of 2.3 $\times$ 10$^{-12}$. \\
\\
$\bullet$ A power-law model gives good fits to each of the PSDs (at lower frequencies) and the slopes range between $1.4$ and $3.1$ but the vast majority are between $1.7$ and $2.8$, so steeper than flicker noise.  No evidence for any QPOs were seen. \\
\\
$\bullet$ An anti-correlation in flux and hardness ratio is found in the long term data which implies a harder when brighter trend.\\
\\
$\bullet$ The flux and spectral analyses indicate that both particle acceleration and synchrotron cooling processes make an important
contribution to the emission from this blazar. 

\section*{Acknowledgements}
\noindent
This research is based on observations obtained with {\it XMM-Newton}, an ESA science mission with instruments and contributions directly funded by ESA member states and NASA. Data from the Steward Observatory spectropolarimetric monitoring project were used. This programme is supported by Fermi Guest Investigator grants NNX08AW56G, NNX09AU10G, NNX12AO93G, and NNX15AU81G. \\
\\
We thank the anonymous referees for useful comments. HG acknowledges financial support from the Department of Science \& Technology (DST), Government of India, through the INSPIRE faculty award IFA17-PH197 at ARIES, Nainital, India. ACG is thankful to Dr. Main Pal for discussion about XMM-Newton data analysis. ML is supported by the China Postdoctoral Science Foundation (Grant No. 2021M693089).

\section*{Data Availability}

The data sets were derived from sources in the public domain: [XMM-Newton, https://heasarc.gsfc.nasa.gov/db-perl/W3Browse/w3browse.pl]. The data underlying this article will be shared on reasonable request to the corresponding author.

\section*{Supporting Information}

Supplementary data are available at MNRAS online. \\
\\
{\bf Figure 1.} All {\it XMM-Newton} light curves in the total energy range 0.2 -- 10 keV of the blazar 3C 273, labelled with its observation IDs in each panel. \\
\\
{\bf Figure 2.} All {\it XMM-Newton} light curves in the soft energy range 0.2 -- 2 keV and 2 -- 10 keV of the blazar 3C 273, labelled with its observation IDs in each panel. \\
\\
{\bf Figure 3.} Hardness ratio plots for all observation IDs of {\it XMM-Newton} of the blazar 3C 273. \\
\\
{\bf Figure 4.} Power spectral density plots of the light curves for all observation IDs of {\it XMM-Newton} of the blazar 3C 273 in the soft energy band. In the each plot the continuous line shows power-law fitting of the red noise and the red dotted line shows a level 3$\sigma$ above the red noise level. The observational ID and power spectral index are given in the each plot. \\
\\
{\bf Figure 5.} Power spectral density plots of the light curves for all observation IDs of {\it XMM-Newton} of the blazar 3C 273 in the hard energy band. In the each plot the continuous line shows power-law fitting of the red noise and the red dotted line shows a level 3$\sigma$ above the red noise level. The observational ID and power spectral index are given in the each plot. \\
\\
{\bf Figure 6.} Power spectral density plots of the light curves for all observation IDs of {\it XMM-Newton} of the blazar 3C 273 in the total energy band. In the each plot the continuous line shows power-law fitting of the red noise and the red dotted line shows a level 3$\sigma$ above the red noise level. The observational ID and power spectral index are given in the each plot. 

\bibliography{ms_Gowtami_accepted}

\begin{thebibliography}{83}
\expandafter\ifx\csname natexlab\endcsname\relax\def\natexlab#1{#1}\fi

\bibitem[{{Abdo} {et~al}\mbox{.}(2010){Abdo}, {Ackermann}, {Ajello}, {Baldini},
  {Ballet}, {Barbiellini}, {Bastieri}, {Bechtol}, {Bellazzini}, {Berenji},
  {Blandford}, {Bloom}, {Bonamente}, {Borgland}, {Bouvier}, {Bregeon}, {Brez},
  {Brigida}, {Bruel}, {Burnett}, {Buson}, {Caliandro}, {Cameron}, {Cannon},
  {Caraveo}, {Carrigan}, {Casandjian}, {Cavazzuti}, {Cecchi}, {{\c{C}}elik},
  {Charles}, {Chekhtman}, {Cheung}, {Chiang}, {Ciprini}, {Claus},
  {Cohen-Tanugi}, {Conrad}, {Costamante}, {Dermer}, {de Angelis}, {de Palma},
  {Silva}, {Drell}, {Dubois}, {Dumora}, {Farnier}, {Favuzzi}, {Fegan}, {Focke},
  {Frailis}, {Fukazawa}, {Funk}, {Fusco}, {Gargano}, {Gasparrini}, {Gehrels},
  {Germani}, {Giglietto}, {Giommi}, {Giordano}, {Glanzman}, {Godfrey},
  {Grenier}, {Grondin}, {Guiriec}, {Hayashida}, {Hays}, {Hill}, {Horan},
  {Hughes}, {J{\'o}hannesson}, {Johnson}, {Johnson}, {Kamae}, {Katagiri},
  {Kataoka}, {Kawai}, {Kn{\"o}dlseder}, {Kuss}, {Lande}, {Larsson},
  {Latronico}, {Lemoine-Goumard}, {Llena Garde}, {Longo}, {Loparco}, {Lott},
  {Lovellette}, {Lubrano}, {Madejski}, {Makeev}, {Mansutti}, {Massaro},
  {Mazziotta}, {McConville}, {McEnery}, {Meurer}, {Michelson}, {Mitthumsiri},
  {Mizuno}, {Moiseev}, {Monte}, {Monzani}, {Morselli}, {Moskalenko}, {Murgia},
  {Nolan}, {Norris}, {Nuss}, {Ohsugi}, {Omodei}, {Orlando}, {Ormes}, {Paneque},
  {Panetta}, {Pelassa}, {Pepe}, {Pesce-Rollins}, {Piron}, {Porter},
  {Rain{\`o}}, {Rando}, {Razzano}, {Reimer}, {Reimer}, {Ritz}, {Rodriguez},
  {Romani}, {Roth}, {Ryde}, {Sadrozinski}, {Sander}, {Scargle}, {Schalk},
  {Sgr{\`o}}, {Siskind}, {Smith}, {Spandre}, {Spinelli}, {Starck}, {Strickman},
  {Suson}, {Tajima}, {Takahashi}, {Takahashi}, {Tanaka}, {Thayer}, {Thayer},
  {Thompson}, {Tibaldo}, {Torres}, {Tosti}, {Tramacere}, {Uchiyama}, {Usher},
  {Vasileiou}, {Vilchez}, {Vitale}, {Waite}, {Wang}, {Wehrle}, {Winer}, {Wood},
  {Yang}, {Ylinen}, \& {Ziegler}}]{2010ApJ...714L..73A}
{Abdo} A.~A. {et~al.}, 2010, \apjl, 714, L73

\bibitem[{{Aggrawal} {et~al}\mbox{.}(2018){Aggrawal}, {Pandey}, {Gupta},
  {Zhang}, {Wiita}, {Yadav}, \& {Tiwari}}]{2018MNRAS.480.4873A}
{Aggrawal} V., {Pandey} A., {Gupta} A.~C., {Zhang} Z., {Wiita} P.~J., {Yadav}
  K.~K., {Tiwari} S.~N., 2018, \mnras, 480, 4873

\bibitem[{{Attridge} {et~al}\mbox{.}(2005){Attridge}, {Wardle}, \&
  {Homan}}]{2005ApJ...633L..85A}
{Attridge} J.~M., {Wardle} J. F.~C., {Homan} D.~C., 2005, \apjl, 633, L85

\bibitem[{{Bhagwan} {et~al}\mbox{.}(2014){Bhagwan}, {Gupta}, {Papadakis}, \&
  {Wiita}}]{2014MNRAS.444.3647B}
{Bhagwan} J., {Gupta} A.~C., {Papadakis} I.~E., {Wiita} P.~J., 2014, \mnras,
  444, 3647

\bibitem[{{Bhagwan} {et~al}\mbox{.}(2016){Bhagwan}, {Gupta}, {Papadakis}, \&
  {Wiita}}]{2016NewA...44...21B}
{Bhagwan} J., {Gupta} A.~C., {Papadakis} I.~E., {Wiita} P.~J., 2016, \na, 44,
  21

\bibitem[{{Bhatta} {et~al}\mbox{.}(2018){Bhatta}, {Mohorian}, \&
  {Bilinsky}}]{2018A&A...619A..93B}
{Bhatta} G., {Mohorian} M., {Bilinsky} I., 2018, \aap, 619, A93

\bibitem[{{Blandford} \& {Rees}(1978)}]{1978PhyS...17..265B}
{Blandford} R.~D., {Rees} M.~J., 1978, \physscr, 17, 265

\bibitem[{{Borse} {et~al}\mbox{.}(2021){Borse}, {Acharya}, {Vaidya},
  {Mukherjee}, {Bodo}, {Rossi}, \& {Mignone}}]{2021A&A...649A.150B}
{Borse} N., {Acharya} S., {Vaidya} B., {Mukherjee} D., {Bodo} G., {Rossi} P.,
  {Mignone} A., 2021, \aap, 649, A150

\bibitem[{{Buisson} {et~al}\mbox{.}(2017){Buisson}, {Lohfink}, {Alston}, \&
  {Fabian}}]{2017MNRAS.464.3194B}
{Buisson} D.~J.~K., {Lohfink} A.~M., {Alston} W.~N., {Fabian} A.~C., 2017,
  \mnras, 464, 3194

\bibitem[{{Burbidge} {et~al}\mbox{.}(1974){Burbidge}, {Jones}, \&
  {O'Dell}}]{1974ApJ...193...43B}
{Burbidge} G.~R., {Jones} T.~W., {O'Dell} S.~L., 1974, \apj, 193, 43

\bibitem[{{Burke} {et~al}\mbox{.}(2021){Burke}, {Shen}, {Blaes}, {Gammie},
  {Horne}, {Jiang}, {Liu}, {McHardy}, {Morgan}, {Scaringi}, \&
  {Yang}}]{2021Sci...373..789B}
{Burke} C.~J. {et~al.}, 2021, Science, 373, 789

\bibitem[{{Calafut} \& {Wiita}(2015)}]{2015JApA...36..255C}
{Calafut} V., {Wiita} P.~J., 2015, Journal of Astrophysics and Astronomy, 36,
  255

\bibitem[{{Carini} {et~al}\mbox{.}(2020){Carini}, {Wehrle}, {Wiita}, {Ward}, \&
  {Pendleton}}]{2020ApJ...903..134C}
{Carini} M., {Wehrle} A.~E., {Wiita} P.~J., {Ward} Z., {Pendleton} K., 2020,
  \apj, 903, 134

\bibitem[{{Chakrabarti} \& {Wiita}(1993)}]{1993ApJ...411..602C}
{Chakrabarti} S.~K., {Wiita} P.~J., 1993, \apj, 411, 602

\bibitem[{{Chernyakova} {et~al}\mbox{.}(2007){Chernyakova}, {Neronov},
  {Courvoisier}, {T{\"u}rler}, {Soldi}, {Beckmann}, {Lubi{\'n}ski}, {Walter},
  {Page}, {Stuhlinger}, {Staubert}, \& {McHardy}}]{2007A&A...465..147C}
{Chernyakova} M. {et~al.}, 2007, \aap, 465, 147

\bibitem[{{Chidiac} {et~al}\mbox{.}(2016){Chidiac}, {Rani}, {Krichbaum},
  {Angelakis}, {Fuhrmann}, {Nestoras}, {Zensus}, {Sievers}, {Ungerechts},
  {Itoh}, {Fukazawa}, {Uemura}, {Sasada}, {Gurwell}, \&
  {Fedorova}}]{2016A&A...590A..61C}
{Chidiac} C. {et~al.}, 2016, \aap, 590, A61

\bibitem[{{Collmar} {et~al}\mbox{.}(2000){Collmar}, {Reimer}, {Bennett},
  {Bloemen}, {Hermsen}, {Lichti}, {Ryan}, {Sch{\"o}nfelder}, {Steinle},
  {Williams}, \& {B{\"o}ttcher}}]{2000A&A...354..513C}
{Collmar} W. {et~al.}, 2000, \aap, 354, 513

\bibitem[{{Courvoisier} {et~al}\mbox{.}(2003){Courvoisier}, {Beckmann},
  {Bourban}, {Chenevez}, {Chernyakova}, {Deluit}, {Favre}, {Grindlay}, {Lund},
  {O'Brien}, {Page}, {Produit}, {T{\"u}rler}, {Turner}, {Staubert},
  {Stuhlinger}, {Walter}, \& {Zdziarski}}]{2003A&A...411L.343C}
{Courvoisier} T.~J.~L. {et~al.}, 2003, \aap, 411, L343

\bibitem[{{Dhiman} {et~al}\mbox{.}(2021){Dhiman}, {Gupta}, {Gaur}, \&
  {Wiita}}]{2021MNRAS.506...1198D}
{Dhiman} V., {Gupta} A.~C., {Gaur} H., {Wiita} P.~J., 2021, \mnras, 506, 1198

\bibitem[{{Edelson} {et~al}\mbox{.}(2002){Edelson}, {Turner}, {Pounds},
  {Vaughan}, {Markowitz}, {Marshall}, {Dobbie}, \& {Warwick}}]{Edelson2002}
{Edelson} R., {Turner} T.~J., {Pounds} K., {Vaughan} S., {Markowitz} A.,
  {Marshall} H., {Dobbie} P., {Warwick} R., 2002, ApJ, 568, 610

\bibitem[{{Edelson} \& {Krolik}(1988)}]{1988ApJ...333..646E}
{Edelson} R.~A., {Krolik} J.~H., 1988, \apj, 333, 646

\bibitem[{{Esposito} {et~al}\mbox{.}(2015){Esposito}, {Walter}, {Jean},
  {Tramacere}, {T{\"u}rler}, {L{\"a}hteenm{\"a}ki}, \&
  {Tornikoski}}]{2015A&A...576A.122E}
{Esposito} V., {Walter} R., {Jean} P., {Tramacere} A., {T{\"u}rler} M.,
  {L{\"a}hteenm{\"a}ki} A., {Tornikoski} M., 2015, \aap, 576, A122

\bibitem[{{Fan} {et~al}\mbox{.}(2014){Fan}, {Kurtanidze}, {Liu}, {Richter},
  {Chanishvili}, \& {Yuan}}]{2014ApJS..213...26F}
{Fan} J.~H., {Kurtanidze} O., {Liu} Y., {Richter} G.~M., {Chanishvili} R.,
  {Yuan} Y.~H., 2014, \apjs, 213, 26

\bibitem[{{Fan} {et~al}\mbox{.}(2009){Fan}, {Peng}, {Tao}, {Qian}, \&
  {Shen}}]{2009AJ....138.1428F}
{Fan} J.~H., {Peng} Q.~S., {Tao} J., {Qian} B.~C., {Shen} Z.~Q., 2009, \aj,
  138, 1428

\bibitem[{{Fernandes} {et~al}\mbox{.}(2020){Fernandes},
  {Pati{\~n}o-{\'A}lvarez}, {Chavushyan}, {Schlegel}, \&
  {Vald{\'e}s}}]{2020MNRAS.497.2066F}
{Fernandes} S., {Pati{\~n}o-{\'A}lvarez} V.~M., {Chavushyan} V., {Schlegel}
  E.~M., {Vald{\'e}s} J.~R., 2020, \mnras, 497, 2066

\bibitem[{{Gaur} {et~al}\mbox{.}(2010){Gaur}, {Gupta}, {Lachowicz}, \&
  {Wiita}}]{2010ApJ...718..279G}
{Gaur} H., {Gupta} A.~C., {Lachowicz} P., {Wiita} P.~J., 2010, \apj, 718, 279

\bibitem[{{Ghisellini} {et~al}\mbox{.}(1997){Ghisellini}, {Villata}, {Raiteri},
  {Bosio}, {de Francesco}, {Latini}, {Maesano}, {Massaro}, {Montagni}, {Nesci},
  {Tosti}, {Fiorucci}, {Pian}, {Maraschi}, {Treves}, {Comastri}, \&
  {Mignoli}}]{1997A&A...327...61G}
{Ghisellini} G. {et~al.}, 1997, \aap, 327, 61

\bibitem[{{Gonz{\'a}lez-Mart{\'\i}n} \& {Vaughan}(2012)}]{Gonzales2012}
{Gonz{\'a}lez-Mart{\'\i}n} O., {Vaughan} S., 2012, A\&A, 544, A80

\bibitem[{{Gopal-Krishna} \& {Wiita}(1992)}]{1992A&A...259..109G}
{Gopal-Krishna}, {Wiita} P.~J., 1992, \aap, 259, 109

\bibitem[{{Grandi} \& {Palumbo}(2004)}]{2004Sci...306..998G}
{Grandi} P., {Palumbo} G. G.~C., 2004, Science, 306, 998

\bibitem[{{Gupta} {et~al}\mbox{.}(2004){Gupta}, {Banerjee}, {Ashok}, \&
  {Joshi}}]{2004A&A...422..505G}
{Gupta} A.~C., {Banerjee} D.~P.~K., {Ashok} N.~M., {Joshi} U.~C., 2004, \aap,
  422, 505

\bibitem[{{Gupta} {et~al}\mbox{.}(2016){Gupta}, {Kalita}, {Gaur}, \&
  {Duorah}}]{2016MNRAS.462.1508G}
{Gupta} A.~C., {Kalita} N., {Gaur} H., {Duorah} K., 2016, \mnras, 462, 1508

\bibitem[{{Hagen-Thorn} {et~al}\mbox{.}(2008){Hagen-Thorn}, {Larionov},
  {Jorstad}, {Arkharov}, {Hagen-Thorn}, {Efimova}, {Larionova}, \&
  {Marscher}}]{2008ApJ...672...40H}
{Hagen-Thorn} V.~A., {Larionov} V.~M., {Jorstad} S.~G., {Arkharov} A.~A.,
  {Hagen-Thorn} E.~I., {Efimova} N.~V., {Larionova} L.~V., {Marscher} A.~P.,
  2008, \apj, 672, 40

\bibitem[{{Hufnagel} \& {Bregman}(1992)}]{1992ApJ...386..473H}
{Hufnagel} B.~R., {Bregman} J.~N., 1992, \apj, 386, 473

\bibitem[{{Impey} {et~al}\mbox{.}(1989){Impey}, {Malkan}, \&
  {Tapia}}]{1989ApJ...347...96I}
{Impey} C.~D., {Malkan} M.~A., {Tapia} S., 1989, \apj, 347, 96

\bibitem[{{Jester} {et~al}\mbox{.}(2006){Jester}, {Harris}, {Marshall}, \&
  {Meisenheimer}}]{2006ApJ...648..900J}
{Jester} S., {Harris} D.~E., {Marshall} H.~L., {Meisenheimer} K., 2006, \apj,
  648, 900

\bibitem[{{Jorstad} {et~al}\mbox{.}(2005){Jorstad}, {Marscher}, {Lister},
  {Stirling}, {Cawthorne}, {Gear}, {G{\'o}mez}, {Stevens}, {Smith}, {Forster},
  \& {Robson}}]{2005AJ....130.1418J}
{Jorstad} S.~G. {et~al.}, 2005, \aj, 130, 1418

\bibitem[{{Jorstad} {et~al}\mbox{.}(2001){Jorstad}, {Marscher}, {Mattox},
  {Wehrle}, {Bloom}, \& {Yurchenko}}]{2001ApJS..134..181J}
{Jorstad} S.~G., {Marscher} A.~P., {Mattox} J.~R., {Wehrle} A.~E., {Bloom}
  S.~D., {Yurchenko} A.~V., 2001, \apjs, 134, 181

\bibitem[{Kalita {et~al}\mbox{.}(2015)Kalita, Gupta, Wiita, Bhagwan, \&
  Duorah}]{Kalita2015}
Kalita N., Gupta A.~C., Wiita P.~J., Bhagwan J., Duorah K., 2015, MNRAS, 451,
  1356

\bibitem[{{Kalita} {et~al}\mbox{.}(2017){Kalita}, {Gupta}, {Wiita}, {Dewangan},
  \& {Duorah}}]{2017MNRAS.469.3824K}
{Kalita} N., {Gupta} A.~C., {Wiita} P.~J., {Dewangan} G.~C., {Duorah} K., 2017,
  \mnras, 469, 3824

\bibitem[{{Kataoka} {et~al}\mbox{.}(2002){Kataoka}, {Tanihata}, {Kawai},
  {Takahara}, {Takahashi}, {Edwards}, \& {Makino}}]{2002MNRAS.336..932K}
{Kataoka} J., {Tanihata} C., {Kawai} N., {Takahara} F., {Takahashi} T.,
  {Edwards} P.~G., {Makino} F., 2002, \mnras, 336, 932

\bibitem[{{Kellermann} {et~al}\mbox{.}(1989){Kellermann}, {Sramek}, {Schmidt},
  {Shaffer}, \& {Green}}]{1989AJ.....98.1195K}
{Kellermann} K.~I., {Sramek} R., {Schmidt} M., {Shaffer} D.~B., {Green} R.,
  1989, \aj, 98, 1195

\bibitem[{{Kirk} {et~al}\mbox{.}(1998){Kirk}, {Rieger}, \&
  {Mastichiadis}}]{1998A&A...333..452K}
{Kirk} J.~G., {Rieger} F.~M., {Mastichiadis} A., 1998, \aap, 333, 452

\bibitem[{{Krawczynski}(2004)}]{2004NewAR..48..367K}
{Krawczynski} H., 2004, \nar, 48, 367

\bibitem[{{Kundu} \& {Gupta}(2014)}]{2014MNRAS.444L..16K}
{Kundu} E., {Gupta} N., 2014, \mnras, 444, L16

\bibitem[{{Lachowicz} {et~al}\mbox{.}(2009){Lachowicz}, {Gupta}, {Gaur}, \&
  {Wiita}}]{2009A&A...506L..17L}
{Lachowicz} P., {Gupta} A.~C., {Gaur} H., {Wiita} P.~J., 2009, \aap, 506, L17

\bibitem[{{Li} {et~al}\mbox{.}(2020){Li}, {Zhang}, {Jin}, {Du}, {Cui}, {Liu},
  \& {Wang}}]{2020ApJ...897...18L}
{Li} Y.-R., {Zhang} Z.-X., {Jin} C., {Du} P., {Cui} L., {Liu} X., {Wang} J.-M.,
  2020, \apj, 897, 18

\bibitem[{{Liu} {et~al}\mbox{.}(2021){Liu}, {Luo}, {Brandt}, {Brotherton},
  {Gallagher}, {Ni}, {Shemmer}, \& {Timlin}}]{2021ApJ...910..103L}
{Liu} H., {Luo} B., {Brandt} W.~N., {Brotherton} M.~S., {Gallagher} S.~C., {Ni}
  Q., {Shemmer} O., {Timlin}, J.~D. I., 2021, \apj, 910, 103

\bibitem[{{Liu} {et~al}\mbox{.}(2019){Liu}, {Feng}, {Xin}, {Bai}, {Li}, \&
  {Wang}}]{2019ApJ...880..155L}
{Liu} H.~T., {Feng} H.~C., {Xin} Y.~X., {Bai} J.~M., {Li} S.~K., {Wang} F.,
  2019, \apj, 880, 155

\bibitem[{{Magdziarz} \& {Zdziarski}(1995)}]{1995MNRAS.273..837M}
{Magdziarz} P., {Zdziarski} A.~A., 1995, \mnras, 273, 837

\bibitem[{{Mangalam} \& {Wiita}(1993)}]{1993ApJ...406..420M}
{Mangalam} A.~V., {Wiita} P.~J., 1993, \apj, 406, 420

\bibitem[{{Marcha} {et~al}\mbox{.}(1996){Marcha}, {Browne}, {Impey}, \&
  {Smith}}]{1996MNRAS.281..425M}
{Marcha} M.~J.~M., {Browne} I.~W.~A., {Impey} C.~D., {Smith} P.~S., 1996,
  \mnras, 281, 425

\bibitem[{{Marscher}(2014)}]{2014ApJ...780...87M}
{Marscher} A.~P., 2014, \apj, 780, 87

\bibitem[{{Marscher} \& {Gear}(1985)}]{1985ApJ...298..114M}
{Marscher} A.~P., {Gear} W.~K., 1985, \apj, 298, 114

\bibitem[{{Marshall} {et~al}\mbox{.}(2001){Marshall}, {Harris}, {Grimes},
  {Drake}, {Fruscione}, {Juda}, {Kraft}, {Mathur}, {Murray}, {Ogle}, {Pease},
  {Schwartz}, {Siemiginowska}, {Vrtilek}, \& {Wargelin}}]{2001ApJ...549L.167M}
{Marshall} H.~L. {et~al.}, 2001, \apjl, 549, L167

\bibitem[{{Mohan} \& {Mangalam}(2015)}]{Mohan2015}
{Mohan} P., {Mangalam} A., 2015, ApJ, 805, 91

\bibitem[{{M{\"u}cke} {et~al}\mbox{.}(2003){M{\"u}cke}, {Protheroe}, {Engel},
  {Rachen}, \& {Stanev}}]{2003APh....18..593M}
{M{\"u}cke} A., {Protheroe} R.~J., {Engel} R., {Rachen} J.~P., {Stanev} T.,
  2003, Astroparticle Physics, 18, 593

\bibitem[{{Padovani}(2017)}]{2017NatAs...1E.194P}
{Padovani} P., 2017, Nature Astronomy, 1, 0194

\bibitem[{{Page} {et~al}\mbox{.}(2004){Page}, {Turner}, {Done}, {O'Brien},
  {Reeves}, {Sembay}, \& {Stuhlinger}}]{2004MNRAS.349...57P}
{Page} K.~L., {Turner} M.~J.~L., {Done} C., {O'Brien} P.~T., {Reeves} J.~N.,
  {Sembay} S., {Stuhlinger} M., 2004, \mnras, 349, 57

\bibitem[{{Paltani} {et~al}\mbox{.}(1998){Paltani}, {Courvoisier}, \&
  {Walter}}]{1998A&A...340...47P}
{Paltani} S., {Courvoisier} T.~J.~L., {Walter} R., 1998, \aap, 340, 47

\bibitem[{{Paltani} \& {T{\"u}rler}(2005)}]{2005A&A...435..811P}
{Paltani} S., {T{\"u}rler} M., 2005, \aap, 435, 811

\bibitem[{{Pandey} {et~al}\mbox{.}(2017){Pandey}, {Gupta}, \&
  {Wiita}}]{Pandey2017}
{Pandey} A., {Gupta} A.~C., {Wiita} P.~J., 2017, ApJ, 841, 123

\bibitem[{{Pandey} {et~al}\mbox{.}(2018){Pandey}, {Gupta}, \&
  {Wiita}}]{2018ApJ...859...49P}
{Pandey} A., {Gupta} A.~C., {Wiita} P.~J., 2018, \apj, 859, 49

\bibitem[{{Pollack} {et~al}\mbox{.}(2016){Pollack}, {Pauls}, \&
  {Wiita}}]{2016ApJ...820...12P}
{Pollack} M., {Pauls} D., {Wiita} P.~J., 2016, \apj, 820, 12

\bibitem[{{Romero} {et~al}\mbox{.}(1999){Romero}, {Cellone}, \&
  {Combi}}]{Romero1999}
{Romero} G.~E., {Cellone} S.~A., {Combi} J.~A., 1999, A\&AS, 135, 477

\bibitem[{{Romero} {et~al}\mbox{.}(2000){Romero}, {Chajet}, {Abraham}, \&
  {Fan}}]{2000A&A...360...57R}
{Romero} G.~E., {Chajet} L., {Abraham} Z., {Fan} J.~H., 2000, \aap, 360, 57

\bibitem[{{R{\"o}ser} {et~al}\mbox{.}(2000){R{\"o}ser}, {Meisenheimer},
  {Neumann}, {Conway}, \& {Perley}}]{2000A&A...360...99R}
{R{\"o}ser} H.~J., {Meisenheimer} K., {Neumann} M., {Conway} R.~G., {Perley}
  R.~A., 2000, \aap, 360, 99

\bibitem[{{Sambruna} {et~al}\mbox{.}(2001){Sambruna}, {Urry}, {Tavecchio},
  {Maraschi}, {Scarpa}, {Chartas}, \& {Muxlow}}]{2001ApJ...549L.161S}
{Sambruna} R.~M., {Urry} C.~M., {Tavecchio} F., {Maraschi} L., {Scarpa} R.,
  {Chartas} G., {Muxlow} T., 2001, \apjl, 549, L161

\bibitem[{{Savolainen} {et~al}\mbox{.}(2006){Savolainen}, {Wiik}, {Valtaoja},
  \& {Tornikoski}}]{2006A&A...446...71S}
{Savolainen} T., {Wiik} K., {Valtaoja} E., {Tornikoski} M., 2006, \aap, 446, 71

\bibitem[{{Schmidt}(1963)}]{1963Natur.197.1040S}
{Schmidt} M., 1963, \nat, 197, 1040

\bibitem[{{Soldi} {et~al}\mbox{.}(2008){Soldi}, {T{\"u}rler}, {Paltani},
  {Aller}, {Aller}, {Burki}, {Chernyakova}, {L{\"a}hteenm{\"a}ki}, {McHardy},
  {Robson}, {Staubert}, {Tornikoski}, {Walter}, \&
  {Courvoisier}}]{2008A&A...486..411S}
{Soldi} S. {et~al.}, 2008, \aap, 486, 411

\bibitem[{{Stocke} {et~al}\mbox{.}(1991){Stocke}, {Case}, {Donahue}, {Shull},
  \& {Snow}}]{1991ApJ...374...72S}
{Stocke} J.~T., {Case} J., {Donahue} M., {Shull} J.~M., {Snow} T.~P., 1991,
  \apj, 374, 72

\bibitem[{{Uchiyama} {et~al}\mbox{.}(2006){Uchiyama}, {Urry}, {Cheung},
  {Jester}, {Van Duyne}, {Coppi}, {Sambruna}, {Takahashi}, {Tavecchio}, \&
  {Maraschi}}]{2006ApJ...648..910U}
{Uchiyama} Y. {et~al.}, 2006, \apj, 648, 910

\bibitem[{{Urry} \& {Padovani}(1995)}]{1995PASP..107..803U}
{Urry} C.~M., {Padovani} P., 1995, \pasp, 107, 803

\bibitem[{{Vaughan}(2010)}]{2010MNRAS.402..307V}
{Vaughan} S., 2010, \mnras, 402, 307

\bibitem[{{Vaughan} {et~al}\mbox{.}(2003){Vaughan}, {F}, {Warwick}, \&
  {Uttley}}]{2003MNRAS.345.1271V}
{Vaughan} S., {F} R., {Warwick} R.~S., {Uttley} P., 2003, \mnras, 345, 1271

\bibitem[{{Wagner} \& {Witzel}(1995)}]{Wagner1995}
{Wagner} S.~J., {Witzel} A., 1995, ARA\&A, 33, 163

\bibitem[{{Wehrle} {et~al}\mbox{.}(2019){Wehrle}, {Carini}, \&
  {Wiita}}]{2019ApJ...877..151W}
{Wehrle} A.~E., {Carini} M., {Wiita} P.~J., 2019, \apj, 877, 151

\bibitem[{{Yaqoob} \& {Serlemitsos}(2000)}]{2000ApJ...544L..95Y}
{Yaqoob} T., {Serlemitsos} P., 2000, \apjl, 544, L95

\bibitem[{{Zhang} \& {Giannios}(2021)}]{2021MNRAS.502.1145Z}
{Zhang} H., {Giannios} D., 2021, \mnras, 502, 1145

\bibitem[{{Zhang} {et~al}\mbox{.}(2002){Zhang}, {Treves}, {Celotti},
  {Chiappetti}, {Fossati}, {Ghisellini}, {Maraschi}, {Pian}, {Tagliaferri}, \&
  {Tavecchio}}]{2002ApJ...572..762Z}
{Zhang} Y.~H. {et~al.}, 2002, \apj, 572, 762

\bibitem[{{Zhang} {et~al}\mbox{.}(2019){Zhang}, {Gupta}, {Gaur}, {Wiita}, {An},
  {Gu}, {Hu}, \& {Xu}}]{2019ApJ...884..125Z}
{Zhang} Z., {Gupta} A.~C., {Gaur} H., {Wiita} P.~J., {An} T., {Gu} M., {Hu} D.,
  {Xu} H., 2019, \apj, 884, 125

\bibitem[{{Zhang} {et~al}\mbox{.}(2021){Zhang}, {Gupta}, {Gaur}, {Wiita}, {An},
  {Lu}, {Fan}, \& {Xu}}]{2021ApJ...909..103Z}
{Zhang} Z., {Gupta} A.~C., {Gaur} H., {Wiita} P.~J., {An} T., {Lu} Y., {Fan}
  S., {Xu} H., 2021, ApJ, 909, 103

\end{thebibliography}
\bibliographystyle{mnras}

\end{document}